\documentclass[journal,final,onecolumn]{IEEEtran}
\usepackage[tight,footnotesize,center]{subfigure}
\usepackage{amsmath,amsfonts,amssymb,bbm}
\usepackage[footnotesize]{caption}
\usepackage{graphicx,xcolor}
\usepackage{multirow,bigdelim,longtable}
\usepackage{amsthm}
\usepackage{hhline}
\usepackage[nobreak]{cite}
\usepackage{tikz,pgfplots}
\usetikzlibrary{intersections,calc,patterns,arrows}
\theoremstyle{definition}
\newtheorem{thm}{Theorem}

\newtheorem*{thm*}{Theorem}
\usepackage{float}
\restylefloat{table}
\newcommand{\mc}[1]{\mathcal{#1}}

\def\typ_eps{\mathcal{T}_\epsilon^{(n)}}

\IEEEoverridecommandlockouts

\begin{document}
\title{Cooperative Binning for\\Semideterministic Channels}
\author{Ritesh Kolte, Ayfer \"{O}zg\"{u}r, Haim Permuter\\
rkolte@stanford.edu, aozgur@stanford.edu, haimp@bgu.ac.il%
\thanks{This work was presented in part at ISIT 2015. R.~Kolte and A.~\"{O}zg\"{u}r are with the Department of Electrical Engineering at Stanford University. H.~Permuter is with the Department of Electrical and Computer Engineering at Ben-Gurion University of the Negev. The work of R. Kolte and A. \"{O}zg\"{u}r was partly supported by a Stanford Graduate Fellowship, by NSF CAREER award 1254786 and by the Center for Science of Information (CSoI), an NSF Science and Technology Center, under grant agreement CCF-0939370. The work of H. Permuter was supported by the Israel Science Foundation (grant no. 684/11) and the ERC starting grant.}}

\maketitle               

\begin{abstract}
The capacity regions of semideterministic multiuser channels, such as the semideterministic relay channel and the multiple access channel with partially cribbing encoders, have been characterized using the idea of partial-decode-forward. However, the requirement to explicitly decode part of the message at intermediate nodes can be restrictive in some settings; for example, when nodes have different side information regarding the state of the channel. In this paper, we generalize this scheme to \emph{cooperative-bin-forward} by building on the observation that explicit recovering of part of the message is not needed to induce cooperation. Instead, encoders can bin their received signals and cooperatively forward the bin index to the decoder. The main advantage of this new scheme is illustrated by considering state-dependent extensions of the aforementioned semideterministic setups. While partial-decode-forward is not applicable in these new setups, cooperative-bin-forward continues to achieve capacity.
\end{abstract}


\section{Introduction}
\begin{figure}[!th]
\centering
\includegraphics[scale=1]{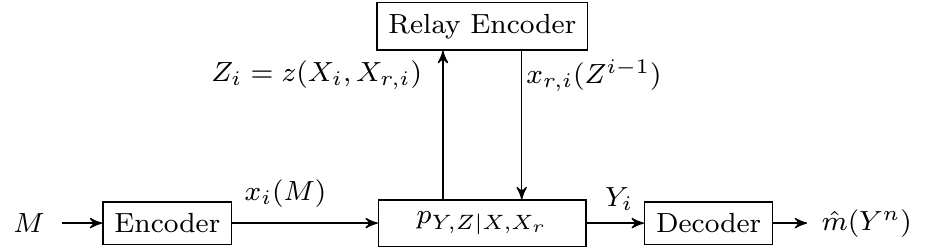}
\caption{Semideterministic Relay Channel
}
\label{fig:semidet_relay}
\end{figure}

The capacity region of the semideterministic relay channel, depicted in Figure~\ref{fig:semidet_relay}, is characterized in \cite{Gam82b} using the partial-decode-forward scheme. In this scheme, the source splits its message into two parts and encodes them using superposition coding. The relay decodes one part of the message, and maps this to a codeword to be transmitted in the next block. The codebooks at the source are generated conditioned on the relay's transmission, which results in coherent transmissions from the source and the relay.

\begin{figure}[!th]
\centering
\includegraphics[scale=1]{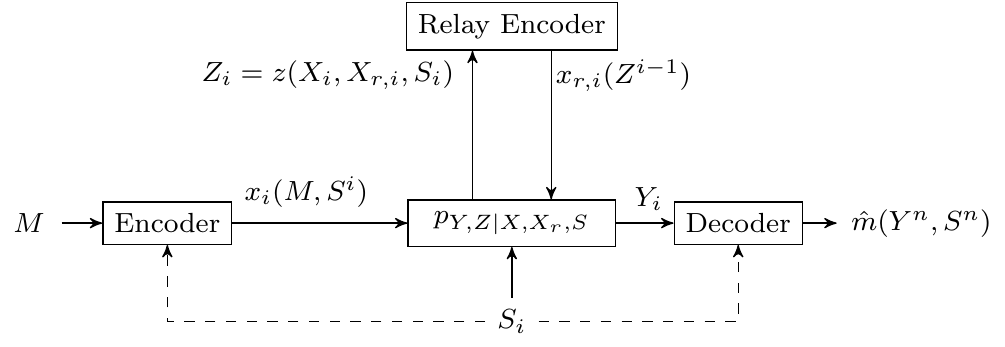}
\caption{State-dependent Semideterministic Relay Channel with Causal State Information at Source and Destination
}
\label{fig:state_semidet_relay}
\end{figure}

Consider now the extension of this model depicted in Figure~\ref{fig:state_semidet_relay}, which corresponds to a state-dependent semideterministic relay channel where the state information is causally available only at the source and the destination. This model captures the natural cellular downlink scenario, in which training enables the source and the destination to learn the channel gain between them (state = channel gain), while a relay could be potentially available to assist the communication, e.g. a wifi access point. In this scenario, it is typically unrealistic to assume that the relay is also able to obtain timely information about the channel state between the source and the destination. In this case, requiring the relay to still decode part of the source message, without any state information, is unduly restrictive and to our knowledge the capacity remains unknown to date. 

The main contribution of this paper is to develop a new scheme which we call \emph{cooperative-bin-forward}. This new scheme does not require the relay to decode part of the message; instead, the relay simply bins its received signal and maps the bin-index to a codeword to be transmitted in the next block. As in partial-decode-forward, the codebooks at the source are generated conditioned on the relay's transmission, resulting in coherent cooperation. This cooperative aspect of the scheme distinguishes it from  bin-forward (a.k.a. hash-forward) that has been considered previously for primitive relay channels in \cite{Kim08}. For the vanilla semideterministic relay channel in Figure~\ref{fig:semidet_relay}, cooperative-bin-forward recovers the capacity achieved by partial-decode-forward. However, while partial-decode-forward is not applicable for the state-dependent semideterministic relay channel in Figure~\ref{fig:state_semidet_relay}, we show that cooperative-bin-forward continues to achieve capacity.

\begin{figure}[!th]
\centering
\includegraphics[scale=1]{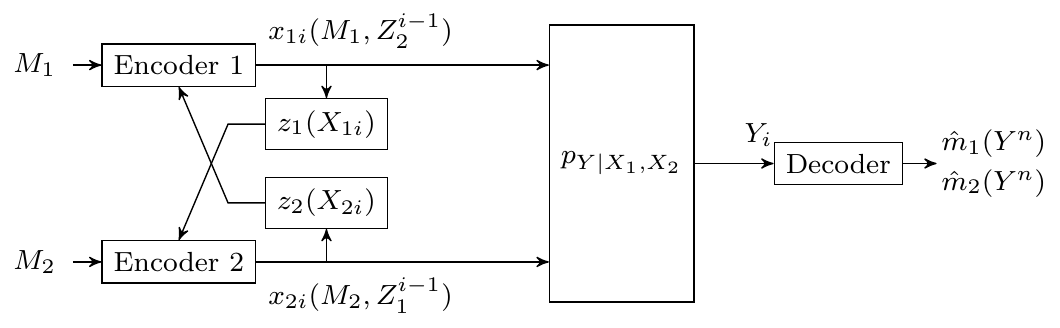}
\caption{Multiple Access Channel with Strictly-Causal Partial-Cribbing Encoders
}
\label{fig:pcrib_mac}
\end{figure}

We next consider another setup where partial-decode-forward is known to be capacity achieving, the multiple-access channel (MAC) with strictly causal partial cribbing encoders, depicted in Figure~\ref{fig:pcrib_mac}. The MAC with cribbing has been introduced by Willems in \cite{Wil85} and its generalization to partial cribbing has been studied in \cite{Asn13}. Compared to the canonical MAC, transmitters here can overhear each other's transmissions while simultaneously transmitting their own data. This possibility, increasingly enabled today by the development of full-duplex radios, is especially appealing since such overheard information can be exploited to induce cooperation among the transmitters by exploiting the natural broadcast nature of the wireless medium without requiring any dedicated resources. Partial cribbing refers to the assumption that the overheard signal is some deterministic function of the signal transmitted by the other transmitter, which allows to capture the signal degradation in the cribbing link via a simple model.\footnote{Earlier work \cite{Asn13} has observed that even if a very coarsely quantized signal is overheard, it can still be sufficient to achieve rates that are close to the best rates achievable with the unrealistic perfect cribbing (overhearing via a noiseless link). Thus, we can in fact manually perform coarse quantization of the overheard signals to simplify operations without a significant loss in performance, while the coarseness simultaneously lends justification to modeling the output of the noisy overhearing channel as a deterministic function of the input.} The MAC with partial cribbing can be regarded as a generalization of the semideterministic relay channel in the sense that when one of the transmitters does not have a message and does not have an outgoing cribbing link, the former reduces to an instance of the latter.

\begin{figure}[!ht]
\centering
\includegraphics[scale=1]{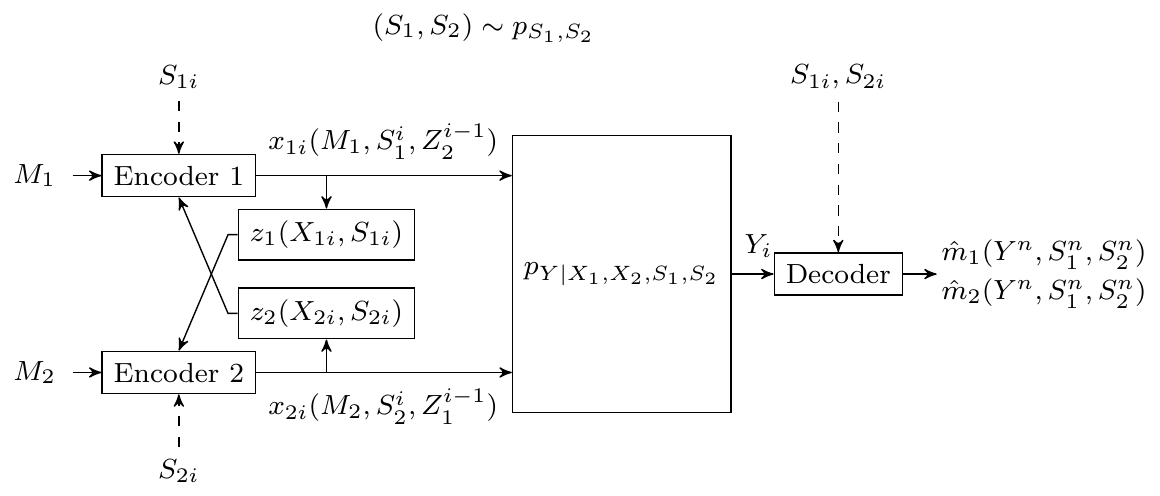}
\caption{State-dependent Multiple-Access Channel with Strictly-Causal Partial-Cribbing Encoders}
\label{fig:state_pcrib_mac}
\end{figure}

As in the semideterministic relay channel, partial-decode-forward achieves the capacity region of the MAC with partial cribbing \cite{Asn13}. Here, we consider a natural extension of this setup with states, depicted in Figure~\ref{fig:state_semidet_relay}.\footnote{This
setup is more general than the setup in Figure~\ref{fig:state_semidet_relay} in the sense that there are two messages and two states, however it is also special in the sense that the partial cribbing links are of the form $z_1(X_{1i},S_{1i})$ and $z_2(X_{2i},S_{2i})$, instead of $z_1(X_{1i},X_{2i},S_{1i})$ and $z_2(X_{1i},X_{2i},S_{2i})$.} The most prevalent example of a situation captured by this model is cellular uplink. In cellular communication, training enables a transmitter to learn the channel gain between itself and the receiver, but assuming knowledge of the channel between any other transmitter and the receiver is unrealistic. Hence, the model in Figure~\ref{fig:state_pcrib_mac} includes a state composed of two components, each known causally only to the corresponding transmitter. These two components are not necessarily independent. Since encoders do not share common state information, partial-decode-forward becomes too restrictive for this setting. Instead, an achievability scheme based on cooperative-bin-forward provides the capacity region.

\begin{figure}[!t]
\centering
\includegraphics[scale=1]{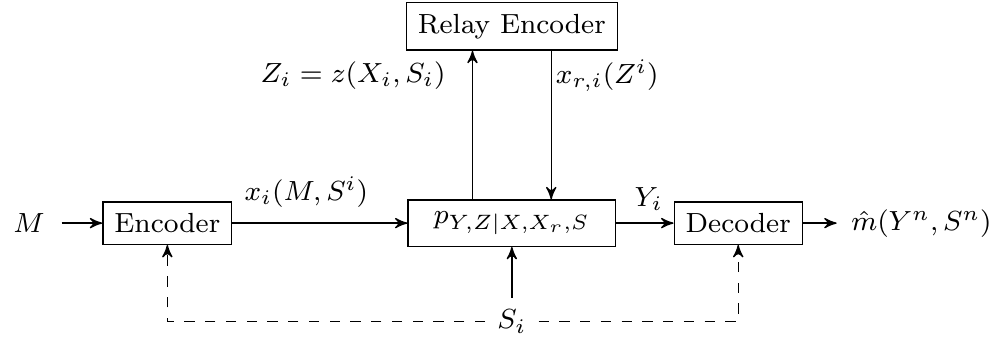}
\caption{State-dependent Semideterministic Relay-Without-Delay Channel with Causal State Information at Source and Destination. For this to make sense, we need to define the received signal at the relay so that it does not depend on the current transmission of the relay, in contrast to Figure~\ref{fig:state_semidet_relay}.
}
\label{fig:delay_state_semidet_relay}
\end{figure}

\begin{figure}[!t]
\centering
\includegraphics[scale=1]{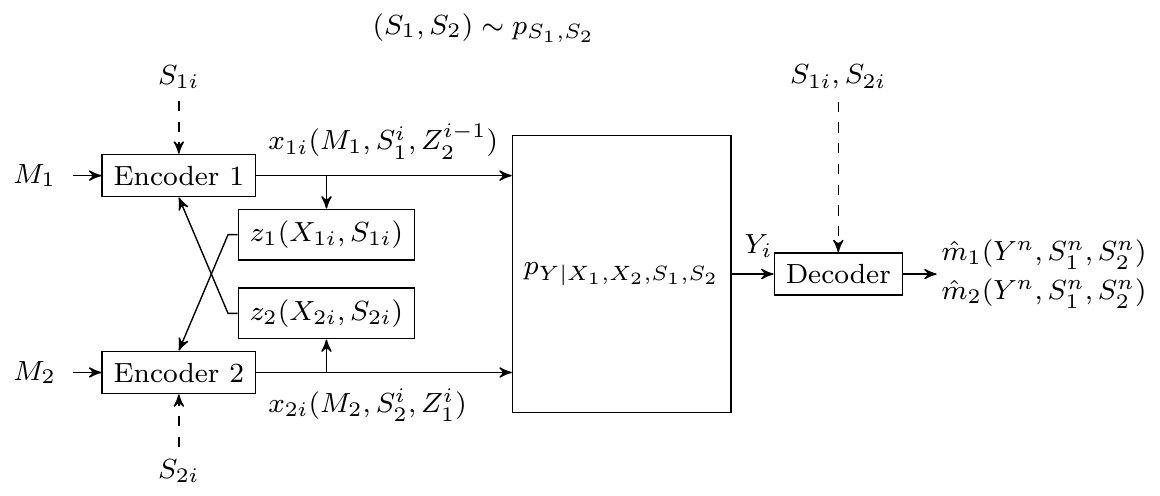}
\caption{State-dependent Multiple-Access Channel with Partially-Cribbing Encoders, one strictly causal and the other causal. Note that both links cannot be changed to causal.}
\label{fig:delay_state_pcrib_mac}
\end{figure}

Finally, motivated by the relay-without-delay channel considered in \cite{Gam07}, we consider ``without-delay'' variations of the two state-dependent setups described above, that are depicted in Figure~\ref{fig:delay_state_semidet_relay} and Figure~\ref{fig:delay_state_pcrib_mac} respectively. In these setups, the strict causality of one of the links is replaced by causality. In the former, which is the state-dependent semideterministic relay-without-delay channel, the transmission of the relay is allowed to depend on its past and current received signal. The capacity region for this setup without state is characterized in \cite{Gam07}, using partial-decode-forward combined with instantaneous relaying (a.k.a. codetrees or Shannon strategies). The latter is a state-dependent multiple access channel with one strictly causal and one causal partial cribbing link. The capacity region for this setup without states is characterized in \cite{Asn13}, again using partial-decode-forward combined with instantaneous relaying. We show that cooperative-bin-forward combined with instantaneous relaying achieves the capacity regions of these setups too, while partial-decode-forward suffers from the same shortcoming encountered for the previous two setups.

\subsection*{Related Work}

We describe here some multiuser setups considered in literature that involve state-dependent channels and/or some form of cooperation, and how they relate to the setups considered in this paper. 

Various cases of state-dependent relay channels have been considered in \cite{Li11,Akh10,Kho08,Kho13,Den13,Zai10,Zai13}. The achievability schemes in these works combine well-known block-Markov relaying ideas such as partial-decode-forward and compress-forward with Shannon strategies or (Gelfand-Pinsker) multicoding. A class of state-dependent orthogonal relay channels with state information only at decoder was considered in \cite{Agu12}, and optimality of a partial-decode-compress-forward scheme was proved. To the best of our knowledge, the state-dependent relay channels considered in this paper have not been previously studied, and as mentioned in the introduction, standard combinations of available ideas are not sufficient to obtain good achievability schemes.

There has been interesting recent work on state-dependent multiple access channels where the state is only known to the encoders. The encoders are not allowed to cooperate in these setups, so the main challenge is to handle the lack of state information at the destination. When the state is known in a strictly causal manner, it was shown in \cite{Lap13a,Lap13b,Li13} that in contrast to point-to-point channels, ignoring the state information is suboptimal. An improvement in achievable rates can be obtained by explicitly communicating the stale state information to the destination. The aforementioned papers accomplished this using block-Markov schemes that encode messages of the current block as well as some information about the state and messages from previous block. When the state information is known noncausally, the work \cite{Phi11} considered the \emph{dirty-paper} special case (additive interference composed of two components each known noncausally to one and only one encoder in Gaussian noise). For this case, a straightforward extension of Gelfand-Pinsker coding turns out to be highly suboptimal. Instead, a structured form of Gelfand-Pinsker coding using lattices is useful for achieving high rates, since it ensures that the overall interference at the destination concentrates on a small set. Finally, the case of common causal state information at the encoders was studied in \cite{Sig05}, which provided an inner bound using Shannon strategies. Recall that the multiple access channels studied in this paper, Figure~\ref{fig:state_pcrib_mac} and Figure~\ref{fig:delay_state_pcrib_mac}, assume also that the destination has full state information, while the encoders cooperate via partial cribbing, thus the main challenge in this paper is to optimally establish cooperation via cribbing among encoders that have disparate state information, rather than handling the lack of state information at the destination.

A few works have considered state-dependent multiple access channels where the state information is available at the encoders as well as the decoder, also with no cooperation between the encoders. It was shown in \cite{Jaf06,Com11,Sen13} that optimal rates can be achieved by effectively treating the state components as time-sharing. These setups can be obtained as a special case of the setup in Figure~\ref{fig:state_pcrib_mac} by setting the partial cribbing links to zero, though the optimal schemes presented previously for these special cases do not provide the necessary insights for establishing cooperation among the encoders. In particular, the causality of cribbing requires the scheme to be block-Markov, and the encoding operation across blocks needs to be such that the two state components can be effectively treated as time-sharing, on top of establishing cooperation inspite of the disparateness of the state components.

Cooperation in multiple access channels was studied by Willems in \cite{Wil83} and \cite{Wil85}, wherein he introduced the notions of conferencing (orthogonal links) and cribbing respectively. In conferencing, dedicated orthogonal links are introduced for cooperation. Cribbing, in contrast, does not assume dedicated resources for cooperation. For example, cribbing can be thought of as exploiting the natural broadcasting nature of the wireless medium for cooperation. Decode-forward based schemes were proved to be optimal for multiple access channels with cribbing. However, the cribbing in \cite{Wil85} was assumed to be perfect (noiseless). To account for the fact that perfect cribbing is unrealistic, the notion of partial cribbing, as described in the introduction, was studied in \cite{Asn13}, and a partial-decode-forward based scheme was shown to be optimal. As mentioned in the introduction, the fact that a part of the message needs to be explicitly decoded in partial-decode-forward, renders its straightforward extension inapplicable for our purpose.

There has also been interest in studying multiple access channels that include states together with some form of cooperation between the encoders, under various assumptions on the state information availability and the form of cooperation \cite{Som08,Bro10,Per11,Zam11,Ema12}. All these works assume that whenever the state information is available, it is available noncausally. The capacity regions for noncausal state information only at one encoder were provided in \cite{Som08} and \cite{Bro10}. The former additionally needed to assume that the informed encoder also knows the other message, while the latter assumed instead that there is a strictly causal or causal perfect cribbing link from the uninformed encoder to the informed encoder. The achievability schemes in both works used Gelfand-Pinsker multicoding at the informed encoder conditioned on the additional information received in the form of message cognition or cribbing. The capacity region for the case of conferencing encoders when noncausal state information is available at all nodes, including the destination, was provided in \cite{Per11} using the idea of double-binning. Achievable rate regions were derived in \cite{Zam11} and \cite{Ema12}, where the former considered perfect cribbing among encoders with noncausal state information only at the encoders, while the latter replaced perfect cribbing by noisy cribbing.

\subsection*{Organization}
The following section describes the models and notation. Section~\ref{sec:mainres} contains the formal statements of the main results described in the introduction. A toy example is considered in Section~\ref{sec:toy} for the purpose of explicitly illustrating the advantage of cooperative-bin-forward over partial-decode-forward. The following sections contain the proofs of the main results. We conclude by describing some open problems in section~\ref{sec:conc}. 

\section{System Models}\label{sec:model}
As standard, capital letters denote random variables, small letters denote realizations, and calligraphic letters denote the alphabet of the corresponding random variable. The notation $\typ_eps$ stands for the $\epsilon$-strongly typical set of sequences for the random variables in context. 

\subsection{State-Dependent Semideterministic Relay Channels}\label{subsec:semidet}
The state-dependent semideterministic relay channel is depicted in Figure~\ref{fig:state_semidet_relay}, and described by the pmf \linebreak $p_{S}(s)p_{Y|X,X_r,S}(y|x,x_r,s)$ and $Z=z(X,X_r,S)$. The encoder and decoder have causal state information. So a $(n,2^{nR},\epsilon)$ code for the above channel consists of the source encoding, relay encoding and decoding functions:
\begin{IEEEeqnarray*}{rCl}
x_{i}  & : & [1:2^{nR}]\times \mc{S}^i \rightarrow \mc{X}, \quad 1\leq i\leq n,\\ 
x_{r,i} & : & \mc{Z}^{i-1} \rightarrow \mc{X}_r, \quad 1\leq i\leq n,\\ 
\hat{m} & : & \mc{Y}^n\times \mc{S}^n \rightarrow [1:2^{nR}],
\end{IEEEeqnarray*}
such that 
$$\text{Pr}\left\{\hat{m}(Y^n,S^n)\neq M\right\} \leq \epsilon,$$
where $M\in[1:2^{nR}]$ denotes the transmitted message. A rate $R$ is said to be \emph{achievable} if for every $\epsilon > 0,$ there exists a $(n,2^{nR},\epsilon)$ code for sufficiently large $n.$ The capacity is defined to be the supremum of  achievable rates. 

The state-dependent semideterministic relay-without-delay channel is depicted in Figure~\ref{fig:delay_state_semidet_relay}, and described by the pmf $p_{S}(s)p_{Y|X,X_r,S}(y|x,x_r,s)$ and $Z = z(X,S)$. The difference from the previous setup is that the relay encoding function is now allowed to depend also on $Z_i$:
\begin{IEEEeqnarray*}{rCl}
x_{r,i} & : &  \mc{Z}^{i} \rightarrow \mc{X}_r, \quad 1\leq i\leq n.
\end{IEEEeqnarray*}
Note that here we need to restrict $Z$ to be $z(X,S)$, instead of $z(X,X_r,S)$.

\subsection{State-Dependent Multiple-Access Channels}
The state-dependent multiple access channel with strictly-causal partial-cribbing encoders is depicted in Figure~\ref{fig:state_pcrib_mac}, and described by the pmf  $p_{S_1,S_2}(s_1,s_2)p_{Y|X_1,X_2,S_1,S_2}(y|x_1,x_2,s_1,s_2)$ and $Z_1=z_1(X_1,S_1)$ and $Z_2=z_2(X_2,S_2)$. The encoders have causal knowledge of the corresponding state components, but no knowledge of the other state component. The decoder is assumed to know both the state components. A $(n,2^{nR_1},2^{nR_2},\epsilon)$ code for the above channel consists of the encoding and decoding functions:
\begin{IEEEeqnarray*}{c}
x_{1,i} : [1:2^{nR_1}]\times \mc{S}_1^i\times \mc{Z}_{2}^{i-1} \rightarrow \mc{X}_1, \quad 1\leq i\leq n,\\ 
x_{2,i} : [1:2^{nR_2}]\times \mc{S}_2^i\times \mc{Z}_{1}^{i-1} \rightarrow \mc{X}_2, \quad 1\leq i\leq n,\\ 
\hat{m}_1: \mc{Y}^n\times \mc{S}_1^n\times \mc{S}_2^n \rightarrow [1:2^{nR_1}],\\
\hat{m}_2: \mc{Y}^n\times \mc{S}_1^n\times \mc{S}_2^n \rightarrow [1:2^{nR_2}],\\
\end{IEEEeqnarray*}
such that 
$$\text{Pr}\left\{\left(\hat{m}_1(Y^n,S_1^n,S_2^n),\hat{m}_2(Y^n,S_1^n,S_2^n)\right)\neq (M_1,M_2)\right\} \leq \epsilon,$$
where $M_1\in[1:2^{nR_1}]$ and $M_2\in[1:2^{nR_2}]$ denote the transmitted messages. A rate pair $(R_1,R_2)$ is said to be \emph{achievable} if for every $\epsilon > 0,$ there exists a $(n,2^{nR_1},2^{nR_2},\epsilon)$ code for sufficiently large $n.$ The capacity region is defined to be the closure of the achievable rate region. 

The ``without-delay'' variation of this setup, also referred to as \emph{causal cribbing}, is depicted in Figure~\ref{fig:delay_state_pcrib_mac}, where one of the partial cribbing links is changed from strictly causal to causal. So, the only difference from the previous setting is that $x_{2i}(M_2,S_2^i,Z_1^{i-1})$ is replaced by $x_{2i}(M_2,S_2^i,Z_1^i)$. 


\section{Main Results}\label{sec:mainres}
In this section, we describe the capacity regions for all the setups described in the previous section. The proofs are presented in subsequent sections. The achievability parts of all the theorems are accomplished by building on the idea of cooperative-bin-forward.

The first result is taken from \cite{Gam82b}. We restate it here and provide a proof of the achievability in Section~\ref{sec:proof_old} using cooperative-bin-forward. Due to the simplicity of the setup, it serves well to bring out the main idea of the new scheme, before we describe results for the more complicated setups. 
\begin{thm}\label{thm:semidet}
The capacity of the semideterministic relay channel, shown in Figure~\ref{fig:semidet_relay}, is given by
\begin{equation}\label{eq:cap_semidet}
C = \max_{p_{X,X_r}(x,x_r)} \min \left\{ I(X,X_r;Y) \, ,\, H(Z|X_r) + I(X;Y|X_r,Z) \right\}.
\end{equation}
\end{thm}

The next result provides an expression for the capacity of the state-dependent semideterministic relay channel. 
\begin{thm}\label{thm:semidet_state}
The capacity of the state-dependent semideterministic relay channel, shown in Figure~\ref{fig:state_semidet_relay}, is given by
\begin{equation}\label{eq:cap_semidet_state}
C = \max_{p_{X_r}(x_r)p_{X|X_r,S}(x|x_r,s)} \min \left\{ I(X,X_r;Y|S) \, ,\, H(Z|S,X_r) + I(X;Y|S,X_r,Z) \right\}.
\end{equation}
\end{thm}

One difference between the capacity expressions of Theorem~\ref{thm:semidet} and Theorem~\ref{thm:semidet_state} is that the mutual information and entropy terms involve a conditioning on $S$. Such an expression would also characterize the capacity if the relay is provided with the state information, and it would be achievable by performing partial-decode-forward while treating the state as a time-sharing sequence. It is quite interesting  then that the capacity expression remains the same even when the relay does not have state information. However, the limitation is reflected in the fact that the choice of pmf is restricted to be $p_{X_r}(x_r)p_{X|X_r,S}(x|x_r,s)$, instead of $p_{X,X_r|S}(x,x_r|s)$. So, the cost of not having state information at the relay is reflected entirely in the limited choice of pmf.

The following theorem states the capacity of the without-delay variation of the above case. The expression involves an auxiliary random variable, which allows the relay to perform instantaneous relaying on top of the binning. This can achieve maximal source-relay cooperation, as conveyed by the following theorem.  
\begin{thm}\label{thm:semidet_state_delay}
The capacity of the state-dependent semideterministic relay-without-delay channel, shown in Figure~\ref{fig:delay_state_semidet_relay}, is given by
\begin{equation}\label{eq:cap_semidet_state_delay}
C = \max_{p_U(u)p_{X|U,S}(x|u,s),X_r=x_r(U,Z)} \min \left\{ I(U,X ; Y|S) , H(Z|U,S) + I(X;Y|U,Z,S) \right\},
\end{equation}
where $|\mc{U}|\leq |\mc{S}|\left(|\mc{X}||\mc{X}_r| - 1\right) + 2$.
\end{thm}

The capacity region for the setup of Theorem~\ref{thm:semidet_state_delay} in the absence of states is characterized in \cite[Proposition~7]{Gam07}. Setting $S$ to be the empty random variable in Theorem~\ref{thm:semidet_state_delay} recovers this result. Note that the objective in \eqref{eq:cap_semidet_state_delay} is the same as that in \eqref{eq:cap_semidet_state} with $X_r$ being replaced by $U$. However, the optimization in \eqref{eq:cap_semidet_state_delay} is over a different domain since the dependence of $X_r$ on $Z$ can now be chosen and is not specified by the channel.

The next two theorems describe the capacity regions for the two multiple-access setups.

\begin{thm}\label{thm:mac}
The capacity region of the state-dependent multiple-access channel with partially cribbing encoders, shown in Figure~\ref{fig:state_pcrib_mac}, is given by the set of rate pairs $(R_1,R_2)$ satisfying
\begin{equation}\label{eq:cap_mac}
\begin{split}
R_1 & \leq I(X_1;Y|U,X_2,Z_1,S_1,S_2) + H(Z_1|U,S_1) ,\\
R_2 & \leq I(X_2;Y|U,X_1,Z_2,S_1,S_2) + H(Z_2|U,S_2),\\
R_1+R_2 & \leq I(X_1,X_2;Y|U,Z_1,Z_2,S_1,S_2)  + H(Z_1,Z_2|U,S_1,S_2),\\
R_1 + R_2 & \leq I(X_1,X_2;Y|S_1,S_2),
\end{split}
\end{equation}
for pmf of the form $$p_U(u)p_{X_1|U,S_1}(x_1 | u,s_1)p_{X_2|U,S_2}(x_2|u,s_2),$$ with $Z_1=z_1(X_1,S_1)$ and $Z_2=z_2(X_2,S_2),$ with ${|\mathcal{U}|\leq \min\{|\mc{S}_1||\mc{S}_2|(|\mc{Y}|-1)+4,|\mc{S}_1||\mc{S}_2|(|\mc{X}_1||\mc{X}_2|-1)+3\} }.$
\end{thm}

\emph{Remark:} It can be shown that the set described in the above theorem is convex, so there is no need to introduce an additional auxiliary random variable for time-sharing. 

As described earlier, the special case of no cribbing (obtained by setting $Z_1=0$ and $Z_2=0$) has been considered in \cite{Jaf06,Com11,Sen13}. For this case, the last inequality becomes redundant and setting the auxiliary random variable $U$ to be the time-sharing random variable in the statement of the above theorem is optimal. The resulting region recovers the results in the aforementioned papers. For the other extreme of constant states, i.e. $S_1=0$ and $S_2=0$, the capacity region in the above theorem recovers the result for strictly causal partial cribbing from \cite{Asn13}.

\begin{thm}\label{thm:mac_delay}
The capacity region of the state-dependent multiple-access channel with partially cribbing encoders in the presence of a causal cribbing link, shown in Figure~\ref{fig:delay_state_pcrib_mac}, is given by the set of rate pairs $(R_1,R_2)$ satisfying
\begin{equation}\label{eq:cap_mac_delay}
\begin{split}
R_1 & \leq I(X_1;Y|U,X_2,Z_1,S_1,S_2) + H(Z_1|U,S_1) ,\\
R_2 & \leq I(X_2;Y|U,X_1,Z_2,S_1,S_2) + H(Z_2|U,S_2,Z_1),\\
R_1+R_2 & \leq I(X_1,X_2;Y|U,Z_1,Z_2,S_1,S_2)  + H(Z_1,Z_2|U,S_1,S_2),\\
R_1 + R_2 & \leq I(X_1,X_2;Y|S_1,S_2),
\end{split}
\end{equation}
for pmf of the form $$p_U(u)p_{X_1|U,S_1}(x_1 | u,s_1)p_{X_2|U,S_2,Z_1}(x_2|u,s_2,z_1),$$ with $Z_1=z_1(X_1,S_1)$ and $Z_2=z_2(X_2,S_2),$ with ${|\mathcal{U}|\leq \min\{|\mc{S}_1||\mc{S}_2|(|\mc{Y}|-1)+4,|\mc{S}_1||\mc{S}_2|(|\mc{X}_1||\mc{X}_2|-1)+3\} }.$
\end{thm}

Note that if $p_{X_2|U,S_2,Z_1}(x_2|u,s_2,z_1)$ in Theorem~\ref{thm:mac_delay} is replaced by $p_{X_2|U,S_2}(x_2|u,s_2)$, then the region becomes identical to that in Theorem~\ref{thm:mac}. Setting $S_1$ and $S_2$ to be constant retrieves the result for causal partial cribbing from \cite{Asn13}.

\section{Illustrative Example}\label{sec:toy}

Consider the following special case of Figure~\ref{fig:state_semidet_relay}.

Let the state $S$ be the ternary random variable
$$p_S(s) = \begin{cases} p/2, & \text{ if } s=0,\\ p/2, & \text{ if } s=1,\\ 1-p,& \text{ if } s=2,\end{cases}$$
where $p<1/2$. The other variables are all binary. The channel $z(X,S)$ is the memory with stuck-at faults channel considered in \cite[Figure~7.7]{Gam12}, while the channel $p_{Y|X,X_r,S}$ is specialized to be a noiseless channel from $X_r$ to $Y$. Formally,
\begin{IEEEeqnarray*}{rCl}
z(X,S) & = & \begin{cases} 0, &\text{ if } S=0,\\ 1, & \text{ if } S=1,\\ X, & \text{ if } S=2,\end{cases}\\
Y & = & X_r.
\end{IEEEeqnarray*}
Recall that the source and the destination know the state information causally while the relay has no state information.

If the relay is required to decode the message, motivated by the optimality of decode-forward in the case of a line network with no state, the achievable rate is limited to be no more than the capacity of the memory with stuck-at faults channel when the state is known causally \emph{only} to the source, which is $1-H_2\left(\frac{p}{2}\right).$ We point out that this cannot be improved by using partial-decode-forward, because the absence of a direct link between the source and destination means that any part of the message that is not decoded by the relay cannot be communicated to the destination in any manner. However, a higher rate can be achieved if the relay simply forwards its received signal, resulting in an effective channel between the source and the destination that is the memory with stuck-at faults channel with state known causally \emph{both} to the source and the destination. The capacity of this channel is $1-p$, which is achieved by multiplexing at the source and demultiplexing at the destination according to the observed state. Thus, a rate $1-p$ which is higher than $1-H_2\left(\frac{p}{2}\right)$ can be achieved. 

What if the channel from the relay to destination is not a noiseless bit-pipe, but a general noisy channel with capacity at least $1-p$? The rate $1-p$ can still be achieved if the operation at the relay is changed from simply forwarding to randomly binning its received signal into $\approx 2^{n(1-p)}$ bins and forwarding a codeword corresponding to the chosen bin. To recover the message, the destination can first decode the bin-index. Since the destination has state information, it can reconstruct the state-multiplexed codebook at the source. Hence, it can recover the message by finding the unique source codeword, if any, that results in the received signal at the relay falling in the correct bin. 

The above example serves to illustrate the limitation of partial-decode-forward when nodes have different side-information. This example did not require cooperative transmissions from the source and the relay, because the source transmission did not directly affect the received signal at the destination. When there is also a direct link between the source and the destination, as allowed in the general models that we consider in this paper, the source and relay need to perform the bin-forward operation in a cooperative fashion.

\section{Proof of Theorem~\ref{thm:semidet}}\label{sec:proof_old}
We demonstrate in this section the achievability of capacity for the semideterministic relay channel using the new scheme cooperative-bin-forward. As described in the introduction, this scheme does not require the relay to decode part of the message. Instead, the relay simply bins its received signal and maps the bin-index to a codeword to be transmitted in the next block. As in partial-decode-forward, the codebooks at the source are generated conditioned on the relay's transmission, resulting in coherent cooperation. The scheme is formally described next.

\subsection*{Proof:} Fix a pmf $p_{X,X_r}(x,x_r)$ and $\epsilon > 0.$ Split $R$ as $R'+R''$, with the message $M$ denoted accordingly as $(M',M'')$. Divide the total communication time into $B$ blocks, each of length $n$.

~

\subsubsection*{Codebook~Generation}

~

For each block $b\in[1:B]$, a codebook is generated independently of the other blocks as follows. 
\begin{itemize}
\item[-] \underline{\emph{Cooperation codewords}}\hfill\\
Generate $2^{n\widetilde{R}}$ codewords $x_{rb}^n(l_{b-1})$ i.i.d. according to $p_{X_r}$, where $l_{b-1}\in[1:2^{n\widetilde{R}}]$. 
\item[-] \underline{\emph{Cribbed codewords}}\footnote{Given the analogy of the source-to-relay link in the relay channel with the cribbing link in the multiple-access channel with cribbing encoders, we call the $z_b^n$ codewords as cribbed codewords.}\hfill\\
For each $l_{b-1}$, generate $2^{nR'}$ codewords $z_b^n(m'_{b}|l_{b-1})$ according to $\prod_{i=1}^np_{Z|X_r}(\cdot|x_{rbi}(l_{b-1}))$, where $m'_b\in[1:2^{nR'}]$. 
\item[-] \underline{\emph{Transmission codewords}}\hfill\\
For each $l_{b-1}$ and each $m'_{b}$, generate $2^{nR''}$ codewords $x_b^n(m''_b|l_{b-1},m'_b)$, where $m''_b\in[1:2^{nR''}]$ according to\\ $\prod_{i=1}^np_{X|X_r,Z}(\cdot|x_{rbi}(l_{b-1}),z_{bi}(m'_{1}|l_{b-1}))$.
\item[-] \underline{\emph{Binning}}\hfill\\
Partition the set of all $\mc{Z}^n$ into $2^{n\widetilde{R}}$ bins, by choosing a bin for each $z^n$ independently and uniformly at random. Denote the index of the chosen bin for $z^n$ by $\textsf{bin}_b(z^n)$.
\end{itemize}

~

\subsubsection*{Encoding}

~

Fix $l_0 = 1$ and $(m'_B,m''_B) = (1,1)$. Since the message in the last block is fixed, the effective rate of communication is $\frac{B-1}{B}R$, which can be made as close as desired to $R$ by choosing a sufficiently large $B$.

In block $b$, $l_{b-1}$ is known to the source encoder. To communicate message $m_b=(m'_b,m''_b)$, it transmits $x_{b}^n(m''_b|l_{b-1},m'_b)$. The relay transmits $x_{rb}^n(l_{b-1})$. Due to the deterministic link from source to relay and the codebook construction, the received signal at the relay in block $b$ is the codeword $z_{b}^n(m'_b|l_{b-1})$. The source and the relay set $l_b$ to be the index of the bin containing $z_{b}^n(m'_b|l_{b-1})$. 

~

\subsubsection*{Decoding}

~

The decoder performs backward decoding, starting from block $B$ and moving towards block $1$, performing the following two steps for each block $b$:
\begin{itemize}
\item[(1)] Assuming that $l_b$ is known from previous operations, the decoder, for each $l_{b-1}\in[1:2^{n\widetilde{R}}]$,  finds the unique $m'_b$ such that  
$$\textsf{bin}_b(z_{b}^n(m'_{b}|l_{b-1})) = l_{b}.$$
Whenever a unique $m'_{b}$ cannot be found for some $l_{b-1}$, the decoder chooses any $m'_{b}$ arbitrarily. So after this operation, the decoder has chosen one $m'_{b}$ for each $l_{b-1}$, given its knowledge of $l_b$. We will signify this explicitly by denoting the chosen message as $\hat{m}'_{b}(l_{b-1},l_{b})$.
\item[(2)] Now the decoder looks for the unique $(\hat{l}_{b-1},\hat{m}''_{b})$ such that 
\begin{equation}\label{eq:dec_semidet}\left(x_{rb}^n(\hat{l}_{b-1})\,,\, z_b^n(\hat{m}'_{b}(\hat{l}_{b-1},l_{b}) | \hat{l}_{b-1})\,,\, x_b^n(\hat{m}''_b | \hat{l}_{b-1},\hat{m}'_{b}(\hat{l}_{b-1},l_{b}))\,,\,y_b^n\right)\in\typ_eps.\end{equation}
\end{itemize}

Note that the first step does not depend on the received signal in block $b$ at the destination. However, it depends on the received signal in block $b+1$, due to the involvement of $l_b$. 

~

\subsubsection*{Probability of Error}

~

The following error analysis reveals that in order to achieve the highest rate, the scheme will set $R' \approx \widetilde{R} \approx H(Z|X_r)$. It is easy to see that when $R' \approx H(Z|X_r)$, given its knowledge of $l_{b-1}$, the relay can indeed recover $m'_b$, even though it is not required to do so in this new scheme. In other words, for a given $l_{b-1}$, since $R' \approx \widetilde{R}$, each message $m'_b$ is mapped to a different bin, and therefore cooperatively communicating the bin index is indeed equivalent to cooperatively communicating the partial message $m'_b$. Thus, cooperative-bin-forward for this basic setup is indeed equivalent to partial-decode-forward. We will see however in the next section that when we have states, even though we still set $R' \approx \widetilde{R}$, the relay will not be able to decode any part of the message so the binning aspect of the scheme will be instrumental.

By symmetry, we can assume without loss of generality that the true messages and bin-indices corresponding to the current block are all 1, i.e. $${(L_{b-1},M'_{b},M''_{b})=(1,1,1)}.$$ We bound the probability of decoding error in block $b$ conditioned on successful decoding for blocks $\{B,B-1,\dots ,b+1\}$, averaged over the randomness in the messages and codebook generation. In particular, successful decoding in block $b+1$ means that $L_{b}$ has been decoded successfully, where we remind ourselves that
$$L_{b} = \textsf{Bin}_b(Z_{b}^n(1|1)).$$

An error occurs in block $b$ only if any of the following events occur:
\begin{itemize}
\item[$(a)$] $\hat{M}'_{b}(1,L_{b})\neq 1$,
\item[$(b)$] $(\hat{L}_{b-1},\hat{M}''_{b}) \neq (1,1)$ given $\hat{M}'_{b}(1,L_{b})= 1.$
\end{itemize}

We analyze the two events in the following. One can notice from the above partitioning of the error events that we are ensuring that $\hat{M}'_{b}(l_{b-1},L_{b})$ is equal to 1 only for $l_{b-1}=L_{b-1}=1$, and not worrying about what $\hat{M}'_{b}(l_{b-1},L_{b})$ is for any other value of $l_{b-1}$. However, it is still important to fix at most one $m'_{b}$ pair for each $l_{b-1}$, even if it is arbitrary for all $l_{b-1}\neq 1$, which is what the first step of the decoding does.\footnote{Of course, one could also discard any $l_{b-1}$ for which a unique $\hat{M}'_{b}(l_{b-1},L_{b})$ cannot be identified; we stick to making one arbitrary choice in such cases only because it makes the exposition simpler.} This allows us to restrict our attention to at most $2^{n(\widetilde{R} + R'')}$ options while analyzing the probability of decoding $L_{b-1}$ incorrectly during the second decoding step, instead of $2^{n(\widetilde{R} + R' + R'')}$. 

~

\subsubsection*{Event $(a)$: $\hat{M}'_{b}(1,L_{b})\neq 1$} 

~

We have
\begin{IEEEeqnarray*}{rCl}
\text{Pr}\left(\hat{M}'_{b}(1,L_{b})\neq 1\right)
&  = & \text{Pr}\left(\textsf{Bin}_b(Z_{b}^n(m'_{b}|1)) = L_{b} \text{ for some } m'_{b}>1\right)\\
& \stackrel{(i)}{=} & \text{Pr}\left(\textsf{Bin}_b(Z_{b}^n(m'_{b}|1)) = \textsf{Bin}_b(Z_{b}^n(1|1)) \text{ for some } m'_{b}>1\right)\\
& \leq & \sum_{m'_{b}>1} \text{Pr}\left(\textsf{Bin}_b(Z_{b}^n(m'_{b}|1)) = \textsf{Bin}_b(Z_{b}^n(1|1))\right)\\
& = &\sum_{m'_{b}>1} \text{Pr}\left(\textsf{Bin}_b(Z_{b}^n(m'_{b}|1)) = \textsf{Bin}_b(Z_{b}^n(1|1)),\, Z_{b}^n(m'_{b}|1) =Z_{b}^n(1|1)\right)\\
&& \quad +\> \sum_{m'_{b}>1} \text{Pr}\left(\textsf{Bin}_b(Z_{b}^n(m'_{b}|1)) = \textsf{Bin}_b(Z_{b}^n(1|1)), \, Z_{b}^n(m'_{b}|1) \neq Z_{b}^n(1|1)\right)\\
& \leq & \sum_{m'_{b}>1} \text{Pr}\left( Z_{b}^n(m'_{b}|1) =Z_{b}^n(1|1)\right)\\
&& \quad +\> \sum_{m'_{b}>1} \text{Pr}\left(\textsf{Bin}_b(Z_{b}^n(m'_{b}|1)) = \textsf{Bin}_b(Z_{b}^n(1|1)) \,\, \big|\,\, Z_{b}^n(m'_{b}|1) \neq Z_{b}^n(1|1)\right)\\
& \stackrel{(ii)}{\leq} & 2^{nR'}\cdot 2^{-n(H(Z|X_r)-\delta(\epsilon))} + 2^{nR'}\cdot 2^{-n\widetilde{R}},
\end{IEEEeqnarray*}
where 
\begin{itemize}
\item[-] $(i)$ follows since $L_{b} = \textsf{Bin}_b(Z_{b}^n(1|1))$,
\item[-] the first term in $(ii)$ is obtained because for $m'_b\neq 1$, the codewords $Z_{b}^n(m'_{b}|1)$ and $Z_{b}^n(1|1)$ are generated independently according to $\prod_{i=1}^np_{Z|X_r}(\cdot | x_{rbi}(l_{b-1}))$, and the second term in $(ii)$ arises because the binning is performed uniformly at random and independently for each sequence,
\end{itemize}
and we use $\delta(\epsilon)$ to denote any function of $\epsilon$ for which $\delta(\epsilon)\rightarrow 0$ as $\epsilon\rightarrow 0$. Hence, we get that
\begin{IEEEeqnarray*}{L}
\text{Pr}\left(\hat{M}'_{b}(1,L_{b})\neq 1\right)\rightarrow 0 \text{ as } n\rightarrow\infty ,
\end{IEEEeqnarray*}
if the following two constraints are satisfied:
\begin{IEEEeqnarray}{rCl}
R' & < & \widetilde{R},\\
R' & < & H(Z|X_r) - \delta(\epsilon).
\end{IEEEeqnarray}

~

\subsubsection*{Event $(b)$: $(\hat{L}_{b-1},\hat{M}''_{b}) \neq (1,1)$ given $\hat{M}'_{b}(1,L_{b})= 1$}

~

The probability of this event is upper bounded by 
\begin{IEEEeqnarray*}{CL}
&\text{Pr}\left(\text{Condition }\eqref{eq:dec_semidet} \text{ is not satisfied by } (l_{b-1},m''_{b}) = (1,1) \,\, \big|\,\, \hat{M}'_{b}(1,L_{b}) = 1\right)\nonumber\\
+ & \text{Pr}\left(\text{Condition }\eqref{eq:dec_semidet}\text{ is satisfied for some } (l_{b-1},m''_{b})\neq (1,1)\,\, \big|\,\, \hat{M}'_{b}(1,L_{b}) = 1 \right).
\end{IEEEeqnarray*}

The first term 
goes to zero as $n\rightarrow\infty$ by the law of large numbers. The second term can be analyzed by standard applications of the packing lemma~\cite{Gam12} as follows: 
\begin{IEEEeqnarray*}{l}
\text{Pr}\left(\text{Condition }\eqref{eq:dec_semidet}\text{ is satisfied for some } (l_{b-1},m''_{b})\neq (1,1)\,\, \big|\,\, \hat{M}'_{b}(1,L_{b}) = 1 \right)\\
\quad \leq \sum_{l_{b-1}=1,m''_b>1} \text{Pr}\left(\left(X_{rb}^n(1), Z_b^n(1 | 1), X_b^n(\hat{m}''_b | 1,1),Y_b^n\right)\in\typ_eps\right)\\
\quad\quad\quad +\> \sum_{l_{b-1}>1,m''_b\geq 1} \text{Pr}\left(\left(X_{rb}^n(\hat{l}_{b-1})\,,\, Z_b^n(\hat{m}'_{b}(\hat{l}_{b-1},L_{b}) | \hat{l}_{b-1})\,,\, X_b^n(\hat{m}''_b | \hat{l}_{b-1},\hat{m}'_{b}(\hat{l}_{b-1},L_{b}))\,,\,Y_b^n\right)\in\typ_eps\right)\\
\quad \leq 2^{nR''}2^{-n(I(X;Y|X_r,Z)-\delta(\epsilon))} + 2^{n(\widetilde{R}+R'')}2^{-n(I(X,X_r,Z;Y)-\delta(\epsilon))},
\end{IEEEeqnarray*}
which follows by applying the packing lemma. Thus, we get that 
$$\text{Pr}\left((\hat{L}_{b-1},\hat{M}''_{b}) \neq (1,1) \,\, \big|\,\, \hat{M}'_{b}(1,L_{b})= 1\right)\rightarrow 0, \text{ as } n\rightarrow\infty,$$
if
\begin{IEEEeqnarray}{rCl}
R'' & < & I(X;Y|X_r,Z) - \delta(\epsilon),\\
\widetilde{R} + R'' & < & I(X,X_r;Y) - \delta(\epsilon).
\end{IEEEeqnarray}

Performing Fourier-Motzkin elimination of $\widetilde{R}$, $R'$ and $R''$ using the rate constraints and also $R=R'+R''$, letting $n\rightarrow \infty$, $B\rightarrow\infty$ and $\epsilon\rightarrow 0$, we get that the rates specified in Theorem~\ref{thm:semidet} are indeed achievable by the scheme presented in this section. One can also simplify the Fourier-Motzkin elimination step by setting $\widetilde{R}$ to be $R'+\delta(\epsilon)$. Since the converse part of this theorem is already known, we do not repeat the arguments here.\hfill\IEEEQED

\section{Proof of Theorem~\ref{thm:semidet_state}}
The achievability part of the theorem is similar to the scheme presented in the previous section.  Due to the availability of causal state information at the source encoder and the decoder, the source encoder constructs codebooks for each state symbol and treats the state sequence as a time-sharing sequence (i.e. it performs multiplexing). Note that since the relay does not have state information, it might not be able to decode part of the message. However, it can still perform the bin-forward operation, allowing us to establish coherence between the source and the relay transmissions without sacrificing unnecessarily on the rate. The decoding is also similar to the previous section, except for the extra demultiplexing component. The converse part of the theorem is presented towards the end of this section. Thus, we see that while partial-decode-forward cannot be applied, cooperative-bin-forward allows us to achieve the capacity region of the state-dependent semideterministic relay channel.

\subsection*{Proof:} Fix a pmf $p_{X_r}(x_r)p_{X|X_r,S}(x|x_r,s)$ and $\epsilon > 0.$ Split $R$ as $R'+R''$, with the  message $M$ denoted accordingly as $(M',M'')$. Divide the total communication time into $B$ blocks, each of length $n$.

\begin{figure}[!th]
\centering
\includegraphics[scale=1.3]{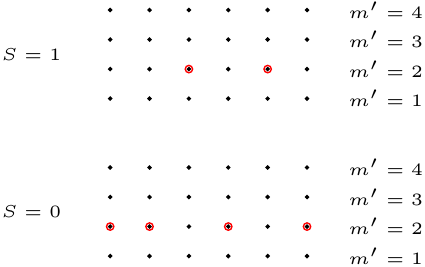}
%
\caption{The figure depicts the cribbed codewords generated for encoding $m'$ for a given $x_r^n(l)$. Each node corresponds to a $z$ symbol that is generated independently according to $p_{Z|X_r,S}(\cdot|x_{r,i}(l),s)$. The red circles show how encoder~1 chooses the codeword if it wants to transmit $m'=2$ and observes $s^n=(0,0,1,0,1,0)$. This construction is not identical but equivalent to that described in \cite[Section 7.4.1]{Gam12}.}
\label{fig:code_a}
\end{figure}

~

\subsubsection*{Codebook~Generation}

~

For each block $b\in[1:B]$, a codebook is generated independently of the other blocks as follows. 
\begin{itemize}
\item[-] \underline{\emph{Cooperation codewords}}\hfill\\
Generate $2^{n\widetilde{R}}$ codewords $x_{rb}^n(l_{b-1})$, i.i.d. according to $p_{X_r}$, where $l_{b-1}\in[1:2^{n\widetilde{R}}]$.
\item[-] \underline{\emph{Cribbed codewords}}\hfill\\
For each $l_{b-1}$ and each $s\in\mc{S}$, generate a codebook of $2^{nR'}$ codewords. The $i$th symbol of such a codeword is chosen independently according to $p_{Z|X_r,S}(\cdot|x_{rbi}(l_{b-1}),s)$. 
The result of this is that for each $l_{b-1}$, each $m'_b\in[1:2^{nR'}]$ and each $s_{b}^n=(s_{b1},s_{b2},\dots ,s_{bn})$, the source encoder can form an effective codeword $z_{b}^n(m'_{b} | l_{b-1},s_{b}^n)$, whose $i$th symbol can be causally chosen as the $i$th symbol of the $m'_{b}$-th codeword from the codebook corresponding to $l_{b-1}$ and $s_{bi}$. See Figure~\ref{fig:code_a}.
\item[-] \underline{\emph{Transmission codewords}}\hfill\\
For each $l_{b-1}$, each $m'_{b}\in[1:2^{nR'}]$ and each $s\in\mc{S}$, generate a codebook of $2^{nR''}$ codewords. The $i$th symbol of such a codeword is generated independently according to $p_{X|X_r,Z,S}(\cdot|x_{rbi}(l_{b-1}),z_{bi}(m'_{b}|l_{b-1},s),s)$. The result of this construction is that for each $l_{b-1}$, each $m'_{b}\in[1:2^{nR'}]$, each $m''_{b}\in[1:2^{nR''}]$ and each $s_{b}^n$, the source encoder can form an effective codeword $x_{b}^n(m''_{b}|l_{b-1},m'_{b},s_{b}^n)$, whose $i$th symbol can be causally chosen as the $i$th symbol of the $m''_{b}$-th codeword from the codebook corresponding to $l_{b-1}$, $m'_{b}$ and $s_{bi}$.
\item[-] \underline{\emph{Binning}}\hfill\\ Partition the set of all $\mc{Z}^n$ into $2^{n\widetilde{R}}$ bins, by choosing a bin for each $z^n$ independently and uniformly at random. Denote the index of the chosen bin for $z^n$ by $\textsf{bin}_b(z^n)$.
\end{itemize}
 
~

\subsubsection*{Encoding}

~

Fix $l_0 = 1$ and $(m'_B,m''_B) = (1,1)$. Since the message in the last block is fixed, the effective rate of communication is $\frac{B-1}{B}R$, which can be made as close as desired to $R$ by choosing a sufficiently large $B$.

In block $b$, $l_{b-1}$ is known to the source encoder. To communicate message $m_b = (m'_b,m''_b)$, it transmits $x_{b}^n(m''_b|l_{b-1},m'_b,s_b^n)$. The relay transmits $x_{rb}^n(l_{b-1})$. Due to the deterministic link from source to relay and the codebook construction, the received signal at the relay in block $b$ is the codeword $z_{b}^n(m'_b|l_{b-1},s_b^n)$. The source and the relay set $l_b$ to be the index of the bin containing $z_{b}^n(m'_b|l_{b-1},s_b^n)$. 

From the encoding operation described above, we can see that the label $l_{b}$ depends on $(l_{b-1},m'_{b},s_{b}^n)$. We do not require the relay to decode $m'_{b}$, but the source and the relay can still establish cooperation by directly performing a binning on the $z_{b}^n$ codeword to agree on the $u_{b+1}^n$ codeword to be used in the next block, thus providing the scheme with the title ``cooperative-bin-forward''. The term \emph{cooperative} is added to emphasize that the source and the relay agree on the binning and transmit coherently. Thus, the scheme achieves cooperation by communicating $l_{b}$ to the relay, instead of $m'_b$. While the relay is not required to decode the partial message, we still need the destination to be able to decode all parts of the transmitted message successfully. In the following, appropriate conditions are imposed so that the destination can utilize the state information at its disposal to achieve successful decoding.

~

\subsubsection*{Decoding}

~

The decoder performs backward decoding, starting from block $B$ and moving towards block $1$, performing the following two steps for each block $b$:
\begin{itemize}
\item[(1)] Assuming that $l_b$ is known from previous operations, the decoder, for each $l_{b-1}\in[1:2^{n\widetilde{R}}]$,  finds the unique $m'_b$ such that  
$$\textsf{bin}_b(z_{b}^n(m'_{b}|l_{b-1},s_b^n)) = l_{b}.$$
Whenever a unique $m'_{b}$ cannot be found for some $l_{b-1}$, the decoder chooses any $m'_{b}$ arbitrarily. So after this operation, the decoder has chosen one $m'_{b}$ for each $l_{b-1}$, given its knowledge of $l_b$ and $s_b^n$. We will signify this explicitly by denoting the chosen message as $\hat{m}'_{b}(l_{b-1},s_b^n,l_{b})$.
\item[(2)] Now the decoder looks for the unique $(\hat{l}_{b-1},\hat{m}''_{b})$ such that 
\begin{equation}\label{eq:dec_semidet_state}\left(x_{rb}^n(\hat{l}_{b-1})\,,\, z_b^n(\hat{m}'_{b}(\hat{l}_{b-1},s_b^n,l_{b}) | \hat{l}_{b-1},s_b^n)\,,\, x_b^n(\hat{m}''_b | \hat{l}_{b-1},\hat{m}'_{b}(\hat{l}_{b-1},s_b^n,l_{b}),s_b^n)\,,\,s_b^n,y_b^n\right)\in\typ_eps.\end{equation}
\end{itemize}

~

\subsubsection*{Probability of Error}

~

In the following error analysis, we will observe that in order to achieve the largest rate the scheme will set $R' \approx H(Z|X_r,S)$. The causal multiplexing-demultiplexing strategy proposed above effectively creates a different codebook for $m'_b$ for each typical $s_b^n$ sequence.  The total number of $z_b^n$ codewords constructed by the source encoder is therefore $\approx 2^{nH(S)}\cdot R' \approx 2^{nH(S,Z|X_r)}$. However, these codewords cannot be distinct since there are only $\approx 2^{nH(Z|X_r)}$ distinct sequences $z_b^n$ (conditioned on $l_{b-1}$). This implies that multiple $(s_b^n, m'_b)$ pairs will be mapped to the same codeword $z_b^n$ and therefore, the relay will be not be able to decode $m'_b$ due to the lack of state information.

By symmetry, we can assume without loss of generality that the true messages and bin-indices corresponding to the current block are all 1, i.e. $${(L_{b-1},M'_{b},M''_{b})=(1,1,1)}.$$ We bound the probability of decoding error in block $b$ conditioned on successful decoding for blocks $\{B,B-1,\dots ,b+1\}$, averaged over the randomness in the messages and codebook generation. In particular, successful decoding in block $b+1$ means that $L_{b}$ has been decoded successfully, where we remind ourselves that
$$L_{b} = \textsf{Bin}_b(Z_{b}^n(1|1,S_b^n)).$$

An error occurs in block $b$ only if any of the following events occur:
\begin{itemize}
\item[$(a)$] $\hat{M}'_{b}(1,S_b^n,L_{b})\neq 1$,
\item[$(b)$] $(\hat{L}_{b-1},\hat{M}''_{b}) \neq (1,1)$ given $\hat{M}'_{b}(1,S_b^n,L_{b})= 1.$
\end{itemize}

We analyze the two terms events in the following two subsections. 

\subsubsection*{Event $(a)$: $\hat{M}'_{b}(1,S_b^n,L_{1,b})\neq 1$} 

We have
\begin{IEEEeqnarray*}{rCl}
\text{Pr}\left(\hat{M}'_{b}(1,S_b^n,L_{1,b})\neq 1\right)
&  = & \text{Pr}\left(\textsf{Bin}_b(Z_{b}^n(m'_{b}|1,S_b^n)) = L_{b} \text{ for some } m'_{b}>1\right)\\
& = & \text{Pr}\left(\textsf{Bin}_b(Z_{b}^n(m'_{b}|1,S_b^n)) = \textsf{Bin}_b(Z_{b}^n(1|1,S_b^n)) \text{ for some } m'_{b}>1\right)\\
& \leq & \sum_{m'_{b}>1} \text{Pr}\left(\textsf{Bin}_b(Z_{b}^n(m'_{b}|1,S_b^n)) = \textsf{Bin}_b(Z_{b}^n(1|1,S_b^n))\right)\\
& = &\sum_{m'_{b}>1} \text{Pr}\left(\textsf{Bin}_b(Z_{b}^n(m'_{b}|1,S_b^n)) = \textsf{Bin}_b(Z_{b}^n(1|1,S_b^n)),\, Z_{b}^n(m'_{b}|1,S_b^n) =Z_{b}^n(1|1,S_b^n)\right)\\
&& \quad +\> \sum_{m'_{b}>1} \text{Pr}\left(\textsf{Bin}_b(Z_{b}^n(m'_{b}|1,S_b^n)) = \textsf{Bin}_b(Z_{b}^n(1|1,S_b^n)), \, Z_{b}^n(m'_{b}|1,S_b^n) \neq Z_{b}^n(1|1,S_b^n)\right)\\
& \leq & \sum_{m'_{b}>1} \text{Pr}\left( Z_{b}^n(m'_{b}|1,S_b^n) =Z_{b}^n(1|1,S_b^n)\right)\\
&& \quad +\> \sum_{m'_{b}>1} \text{Pr}\left(\textsf{Bin}_b(Z_{b}^n(m'_{b}|1,S_b^n)) = \textsf{Bin}_b(Z_{b}^n(1|1,S_b^n)) \, | \, Z_{b}^n(m'_{b}|1,S_b^n) \neq Z_{b}^n(1|1,S_b^n)\right)\\
& \leq & 2^{nR'}\cdot 2^{-n(H(Z|X_r,S)-\delta(\epsilon))} + 2^{nR'}\cdot 2^{-n\widetilde{R}},
\end{IEEEeqnarray*}
where we use $\delta(\epsilon)$ to denote any function of $\epsilon$ for which $\delta(\epsilon)\rightarrow 0$ as $\epsilon\rightarrow 0$. Hence, we get that
\begin{IEEEeqnarray*}{L}
\text{Pr}\left(\hat{M}'_{b}(1,S_b^n,L_{b})\neq 1\right)\rightarrow 0 \text{ as } n\rightarrow\infty ,
\end{IEEEeqnarray*}
if the following two constraints are satisfied:
\begin{IEEEeqnarray}{rCl}
R' & < & \widetilde{R},\\
R' & < & H(Z|X_r,S) - \delta(\epsilon).
\end{IEEEeqnarray}

\subsubsection*{Event $(b)$: $(\hat{L}_{b-1},\hat{M}''_{b}) \neq (1,1)$ given $\hat{M}'_{b}(1,S_b^n,L_{b})= 1$}

The probability of this event is upper bounded by 
\begin{IEEEeqnarray*}{CL}
&\text{Pr}\left(\text{Condition }\eqref{eq:dec_semidet_state} \text{ is not satisfied by } (l_{b-1},m''_{b}) = (1,1) \,\, \big|\,\, \hat{M}'_{b}(1,S_b^n,L_{b}) = 1\right)\nonumber\\
+ & \text{Pr}\left(\text{Condition }\eqref{eq:dec_semidet_state}\text{ is satisfied for some } (l_{b-1},m''_{b})\neq (1,1)\,\, \big|\,\, \hat{M}'_{b}(1,S_b^n,L_{b}) = 1 \right).
\end{IEEEeqnarray*}

The first term 
goes to zero as $n\rightarrow\infty$ by the law of large numbers. The second term can be analyzed by standard applications of the packing lemma~\cite{Gam12} as follows: 
\begin{IEEEeqnarray*}{l}
\text{Pr}\left(\text{Condition }\eqref{eq:dec_semidet_state}\text{ is satisfied for some } (l_{b-1},m''_{b})\neq (1,1)\,\, \big|\,\, \hat{M}'_{b}(1,S_b^n,L_{b}) = 1 \right)\\
\leq \sum_{l_{b-1}=1,m''_b>1} \text{Pr}\left(\left(X_{rb}^n(1), Z_b^n(1 | 1,S_b^n), X_b^n(\hat{m}''_b | 1,1,S_b^n),S_b^n,Y_b^n\right)\in\typ_eps\right)\\
\quad +\> \sum_{l_{b-1}>1,m''_b\geq 1} \text{Pr}\left(\left(X_{rb}^n(\hat{l}_{b-1})\,,\, Z_b^n(\hat{m}'_{b}(\hat{l}_{b-1},S_b^n,L_{b}) | \hat{l}_{b-1},S_b^n)\,,\, X_b^n(\hat{m}''_b | \hat{l}_{b-1},\hat{m}'_{b}(\hat{l}_{b-1},S_b^n,l_{b}),S_b^n)\,,\,S_b^n,Y_b^n\right)\in\typ_eps\right)\\
\leq 2^{nR''}2^{-n(I(X;Y|X_r,Z,S)-\delta(\epsilon))} + 2^{n(\widetilde{R}+R'')}2^{-n(I(X,X_r,Z;Y|S)-\delta(\epsilon))},
\end{IEEEeqnarray*}
which follows by applying the packing lemma. Note that when $l_{b-1}>1$, it so happens due to the codebook construction that $y_b^n$ is independent of all the other sequences for any value of $(m'_{b},m''_{b})$. So the joint distribution of the sequences has the same factorization no matter what $m'_{b}$ is chosen for $l_{b-1}>1$. The only fact that matters for our analysis is that at most one $m'_{b}$ has been chosen somehow for each $l_{b-1}>1$. This allows us to write the fourth event as the union of at most $2^{n(\widetilde{R} + R'')}$ events, where each corresponds to a different value of $(l_{b-1},m''_{b})$.

Thus, we get that 
$$\text{Pr}\left((\hat{L}_{b-1},\hat{M}''_{b}) \neq (1,1) | \hat{M}'_{b}(1,L_{b})= 1\right)\rightarrow 0, \text{ as } n\rightarrow\infty,$$
if
\begin{IEEEeqnarray}{rCl}
R'' & < & I(X;Y|X_r,Z,S) - \delta(\epsilon),\\
\widetilde{R} + R'' & < & I(X,X_r;Y|S) - \delta(\epsilon).
\end{IEEEeqnarray}

Performing Fourier-Motzkin elimination, letting $n\rightarrow \infty$, $B\rightarrow\infty$ and $\epsilon\rightarrow 0$, we get that the rates specified in Theorem~\ref{thm:semidet_state} are indeed achieved by the achievability scheme presented in this section. 

~

\subsubsection*{Converse}

~

Given a reliable code, we have by Fano's inequality
$$H(M|Y^n,S^n) \leq n\epsilon_n,$$
where $\epsilon_n\rightarrow 0$ as $n\rightarrow\infty.$ Then, we prove the first bound on $R$ as follows:
\begin{IEEEeqnarray*}{rCl}
nR & = & H(M)\\
& \stackrel{(a)}{=} & H(M|S^n)\\
& \stackrel{(b)}{=} & H(M,Z^n|S^n)\\
& = & H(Z^n|S^n) + H(M|Z^n,S^n)\\
& \stackrel{(c)}{\leq} & H(Z^n|S^n) + I(M;Y^n|Z^n,S^n) + n\epsilon_n\\
& \stackrel{(d)}{=} & \sum_{i=1}^n H(Z_i|S^n,Z^{i-1}) + \sum_{i=1}^n I(M,X_i;Y_i|Y^{i-1},Z^n,S^n) + n\epsilon_n\\
& \stackrel{(e)}{=} & \sum_{i=1}^n H(Z_i|S^n,Z^{i-1},X_{r,i}) + \sum_{i=1}^n I(M,X_i;Y_i|Y^{i-1},Z^n,S^n,X_r^n) + n\epsilon_n\\
& \stackrel{(f)}{\leq} & \sum_{i=1}^n H(Z_i|S_i,X_{r,i}) + \sum_{i=1}^n I(X_i;Y_i|Z_i,S_i,X_{r,i}) + n\epsilon_n\\
& = & nH(Z_Q|S_Q,X_{rQ},Q) + nI(X_Q;Y_Q|Z_Q,S_Q,X_{rQ},Q) +n\epsilon_n\\
& \leq & nH(Z_Q|S_Q,X_{rQ}) + nI(X_Q;Y_Q|Z_Q,S_Q,X_{rQ}) +n\epsilon_n,
\end{IEEEeqnarray*}
where
\begin{itemize}
\item[-] $Q$ is a random variable uniformly distributed over $[1:n]$, independent of $(X^n,X_r^n,S^n,Y^n)$,
\item[-] $(a)$ follows because $M$ is independent of $S^n$,
\item[-] $(b)$ follows because $Z^n$ is a function of $M$ and $S^n$,
\item[-] $(c)$ follows by Fano's inequality,
\item[-] $(d)$ follows by the chain rule of mutual information and because $X_i$ is a function of $(M,S^n)$,
\item[-] $(e)$ follows because $X_{r,i}$ is a function of $Z^{i-1}$,
\item[-] $(f)$ follows because conditioning reduces entropy and $Y_i$ is independent of other random variables given $(X_i,X_{r,i},S_i)$, and
\item[-] the final step follows because conditioning reduces entropy and $Q-(X_Q,X_{rQ},S_Q)-Y_Q$.
\end{itemize}

The second bound on $R$ is proved as follows:
\begin{IEEEeqnarray*}{rCl}
nR & = & H(M) \\ 
& = & H(M|S^n)\\
& \leq & I(M;Y^n|S^n) + n\epsilon_n\\
& = & I(M,X^n,X_r^n;Y^n|S^n) + n\epsilon_n\\
& \leq & \sum_{i=1}^n I(X_i,X_{r,i};Y_i|S_i) + n\epsilon_n\\
& = & nI(X_Q,X_{rQ};Y_Q|S_Q,Q) + n\epsilon_n\\
& \leq & nI(X_Q,X_{rQ};Y_Q|S_Q) + n\epsilon_n.
\end{IEEEeqnarray*}

Thus, we have
$$R \leq \min (I(X_Q,X_{rQ};Y_Q|S_Q), H(Z_Q|S_Q,X_{rQ}) + I(X_Q;Y_Q|Z_Q,S_Q,X_{rQ})) + \epsilon_n.$$

Note that
\begin{itemize} 
\item[-] $S_Q$ is independent of $Q$ and has marginal pmf $p_S$ due to the i.i.d. assumption on the state;
\item[-] since $S_i$ is independent of $X_{r,i}=x_{r,i}(Z^{i-1})$ for all $1\leq i\leq n$, we have that $S_Q$ is independent of $X_{rQ}$;
\item[-] we have that $p_{Y_Q|X_Q,X_{rQ},S_Q}(y|x,x_{r},s)$ is equal to $p_{Y|X,X_{r},S}(y|x,x_{r},s)$, since $Y_Q$ is the output of the channel when the inputs are $(X_Q,X_{rQ},S_Q)$, and
\item[-] similarly, we also have $Z_Q = z(X_Q,X_{rQ},S_Q).$
\end{itemize}
Hence the joint pmf of the random variables $(X_Q,X_{rQ},S_Q,Y_Q)$ factorizes as
\begin{IEEEeqnarray*}{l}
p_{S_Q,X_Q,X_{rQ},Y_Q}(s,x,x_{r},y)\\
\quad\quad = p_{S_Q}(s)p_{X_{rQ}}(x_{r})p_{X_Q|X_{rQ},S_Q}(x|x_{r},s)p_{Y_Q|X_Q,X_{rQ},S_Q}(y|x,x_{r},s)\\
\quad\quad =  p_{S}(s)p_{X_{rQ}}(x_{r})p_{X_Q|X_{rQ},S_Q}(x|x_{r},s)p_{Y|X,X_{r},S}(y|x,x_{r},s).
\end{IEEEeqnarray*}

So, we can define the random variables $X\triangleq X_Q$, $X_r \triangleq X_{rQ}$, $S\triangleq S_Q$, $Z\triangleq Z_Q$ and $Y\triangleq Y_Q$ to get 
$$R \leq \min \{I(X,X_{r};Y|S), H(Z|S,X_{r}) + I(X;Y|Z,S,X_{r})\} + \epsilon_n,$$
where the pmf of the random variables has the form $p_S(s)p_{X_r}(x_r)p_{X|X_r,S}(x|x_r,s)p_{Y|X,X_r,S}(y|x,x_r,s)$ and $Z=z(X,X_r,S)$. Since $\epsilon_n\rightarrow 0$ as $n\rightarrow\infty$, the converse is proved. 

This concludes the proof of Theorem~\ref{thm:semidet_state}.\hfill\IEEEQED

\section{Proof of Theorem~\ref{thm:semidet_state_delay}}
The achievability part of this theorem is obtained by combining the cooperative-bin-forward scheme from the previous section with instantaneous relaying. This requires an auxiliary random variable, as described next. 

\subsection*{Proof:}
Fix $p_U(u)p_{X|U,S}(x|u,s)$, $X_r = x_r(u,z)$ and $\epsilon>0$. Split $R$ as $R'+R''$, with the message $m$ denoted accordingly as $M=(M',M'')$. Divide the total communication time into $B$ blocks, each of length $n$.

~

\subsubsection*{Codebook~Generation}

~

For each block $b\in[1:B]$, a codebook is generated independently of the other blocks as follows. 
\begin{itemize}
\item[-] \underline{\emph{Cooperation codewords}}\hfill\\
Generate $2^{n\widetilde{R}}$ codewords $u_b^n(l_{b-1})$, i.i.d. according to $p_U$, where $l_{b-1}\in[1:2^{n\widetilde{R}}]$.
\item[-] \underline{\emph{Cribbed codewords}}\hfill\\
For each $l_{b-1}$ and each $s\in\mc{S}$, generate a codebook of $2^{nR'}$ codewords. The $i$th symbol of such a codeword is chosen independently according to $p_{Z|U,S}(\cdot|u_{bi}(l_{b-1}),s)$. 
The result of this is that for each $l_{b-1}$, each $m'_b\in[1:2^{nR'}]$ and each $s_{b}^n=(s_{b1},s_{b2},\dots ,s_{bn})$, the source encoder can form an effective codeword $z_{b}^n(m'_{b} | l_{b-1},s_{b}^n)$, whose $i$th symbol can be causally chosen as the $i$th symbol of the $m'_{b}$-th codeword from the codebook corresponding to $l_{b-1}$ and $s_{bi}$. 
\item[-] \underline{\emph{Transmission codewords}}\hfill\\
For each $l_{b-1}$, each $m'_{b}\in[1:2^{nR'}]$ and each $s\in\mc{S}$, generate a codebook of $2^{nR''}$ codewords. The $i$th symbol of such a codeword is generated independently according to $p_{X|U,Z,S}(\cdot|u_{bi}(l_{b-1}),z_{bi}(m'_{b}|l_{b-1},s),s)$. The result of this construction is that for each $l_{b-1}$, each $m'_{b}\in[1:2^{nR'}]$, each $m''_{b}\in[1:2^{nR''}]$ and each $s_{b}^n$, the source encoder can form an effective codeword $x_{b}^n(m''_{b}|l_{b-1},m'_{b},s_{b}^n)$, whose $i$th symbol can be causally chosen as the $i$th symbol of the $m''_{b}$-th codeword from the codebook corresponding to $l_{b-1}$, $m'_{b}$ and $s_{bi}$.
\item[-] \underline{\emph{Binning}}\hfill\\ Partition the set of all $\mc{Z}^n$ into $2^{n\widetilde{R}}$ bins, by choosing a bin for each $z^n$ independently and uniformly at random. Denote the index of the chosen bin for $z^n$ by $\textsf{bin}_b(z^n)$.
\end{itemize}

~

\subsubsection*{Encoding}

~

Fix $l_0 = 1$ and $(m'_B,m''_B) = (1,1)$. Since the message in the last block is fixed, the effective rate of communication is $\frac{B-1}{B}R$, which can be made as close as desired to $R$ by choosing a sufficiently large $B$.

In block $b$, assuming $l_{b-1}$ is known to the source encoder, it transmits $x_{b}^n(m''_b|l_{b-1},m'_b,s_b^n)$. The relay transmits $x_{rb}^n$, the $i$th symbol of which is obtained as $x_r(u_{bi}(l_{b-1}),z_{bi}(m'_b|l_{b-1},s_b^n))$. At the end of block $b$, the source and the relay set $l_b$ to be the index of the bin containing $z_{b}^n(m'_b|l_{b-1},s_b^n)$. 

~

\subsubsection*{Decoding}

~

The decoding operation is nearly the same as that in the previous section. The decoder performs the following two steps for each block $b$, where $b\in\{B,B-1,\cdots, 1\}$:
\begin{itemize}
\item[(1)] Assuming that $l_b$ is known from previous operations, the decoder, for each $l_{b-1}\in[1:2^{n\widetilde{R}}]$,  finds the unique $m'_b$ such that  
$$\textsf{bin}_b(z_{b}^n(m'_{b}|l_{b-1},s_b^n)) = l_{b}.$$
Whenever a unique $m'_{b}$ cannot be found for some $l_{b-1}$, the decoder chooses any $m'_{b}$ arbitrarily. So after this operation, the decoder has chosen one $m'_{b}$ for each $l_{b-1}$, given its knowledge of $l_b$ and $s_b^n$. We will signify this explicitly by denoting the chosen message as $\hat{m}'_{b}(l_{b-1},s_b^n,l_{b})$.
\item[(2)] Now the decoder looks for the unique $(\hat{l}_{b-1},\hat{m}''_{b})$ such that 
\begin{equation}\label{eq:dec_semidet_state_delay}\left(u_{b}^n(\hat{l}_{b-1})\,,\, z_b^n(\hat{m}'_{b}(\hat{l}_{b-1},s_b^n,l_{b}) | \hat{l}_{b-1},s_b^n)\,,\, x_b^n(\hat{m}''_b | \hat{l}_{b-1},\hat{m}'_{b}(\hat{l}_{b-1},s_b^n,l_{b}),s_b^n)\,,\,s_b^n,y_b^n\right)\in\typ_eps.\end{equation}
\end{itemize}

~

\subsubsection*{Probability of Error}

~

By following a similar path as the previous section, we get the following conditions for vanishing probability of error:
\begin{IEEEeqnarray*}{rCl}
R' & < & \widetilde{R},\\
R' & < & H(Z|U,S) - \delta(\epsilon),\\
R'' & < & I(X;Y|U,Z,S) - \delta(\epsilon),\\
\widetilde{R} + R'' & < & I(U,X;Y|S) - \delta(\epsilon).
\end{IEEEeqnarray*}

Performing Fourier-Motzkin elimination completes the proof of achievability.

~

\subsubsection*{Converse}

~

Given a reliable code, define for each $1\leq i\leq n$, the random variable $U_i\triangleq (Z^{i-1},S^{i-1})$. Note that with this definition, $X_{r,i}$ becomes a function of $(U_i,Z_i)$. We have for any reliable code, by Fano's inequality,
$$H(M|Y^n,S^n) \leq n\epsilon_n.$$

Then,
\begin{IEEEeqnarray*}{rCl}
nR & = & H(M|S^n)\\
& = & H(M,Z^n|S^n)\\
& \leq & H(Z^n|S^n) + I(M;Y^n|S^n,Z^n) + n\epsilon_n\\
& \leq & \sum_{i=1}^n H(Z_i|Z^{i-1},S^{i-1},S_i) + \sum_{i=1}^n I(X_i;Y_i|Z^{i-1},S^{i-1},Z_i,S_i) +n\epsilon_n,
\end{IEEEeqnarray*}
where the final step uses the fact that $Y_i$ is independent of other random variables given $X_i,Z^{i-1},Z_i,S_i$, since $X_{r,i}$ is a function of $Z^i$. Using the definition of $U_i$ in the above, we get that
\begin{IEEEeqnarray*}{rCl}
nR & \leq & \sum_{i=1}^n H(Z_i|U_i,S_i) + \sum_{i=1}^n I(X_i;Y_i|U_i,Z_i,S_i) +n\epsilon_n\\
& = & nH(Z|U_Q,S_Q,Q) + nI(X_Q;Y_Q|U_Q,Z_Q,S_Q,Q) + n\epsilon_n,
\end{IEEEeqnarray*}
where $Q$ is uniformly distributed over $[1:n]$ and independent of $U^n,X^n,X_r^n,S^n,Y^n.$


The remaining bound on $R$ is proved below: 
\begin{IEEEeqnarray*}{rCl}
nR & = & H(M|S^n)\\
& \leq & I(M;Y^n|S^n) + n\epsilon_n\\
& = & \sum_{i=1}^n I(M ; Y_i | Y^{i-1}, S^n) + n\epsilon_n\\
& \stackrel{(a)}{=} & \sum_{i=1}^n I(M,X_i,X_{r,i},Z^{i-1},S^{i-1} ; Y_i | Y^{i-1}, S^n) + n\epsilon_n\\
& \leq & \sum_{i=1}^n I(X_i,X_{r,i},Z^{i-1},S^{i-1} ; Y_i | S_i) + n\epsilon_n\\
& = & \sum_{i=1}^n I(U_i,X_i,X_{r,i} ; Y_i | S_i) + n\epsilon_n\\
& = & \sum_{i=1}^n I(U_i,X_i ; Y_i | S_i) + n\epsilon_n\\
& = & nI(U_Q,X_Q;Y_Q|S_Q,Q) + n\epsilon_n\\
& \leq & nI(Q,U_Q,X_Q;Y_Q|S_Q)+ n\epsilon_n,
\end{IEEEeqnarray*}
where step $(a)$ is true since $(X_i,X_{r,i},Z^{i-1},S^{i-1})$ is a function of $(M,S^n)$, and step $(b)$ follows because $X_{r,i}$ is a function of $(U_i,Z_i)$, hence a function of $(U_i,X_i,S_i).$ 

Following similar arguments as the previous section, we can define $U\triangleq (Q,U_Q)$, $X\triangleq X_Q$, $X_r \triangleq X_{rQ}$, $S\triangleq S_Q$, $Z\triangleq Z_Q$ and $Y\triangleq Y_Q$ to get 
$$R\leq \min\{I(U,X;Y|S), H(Z|U,S)+I(X;Y|U,Z,S)\} +\epsilon_n,$$
where the pmf of the random variables has the form $p_S(s)p_U(u)p_{X|U,S}(x|u,s)p_{Y|X,X_r,S}(y|x,x_r,s)$, $Z=z(X,S)$ and $X_r=x_r(U,Z)$. Since $\epsilon_n\rightarrow 0$ as $n\rightarrow 0$, the converse is completed. The bound on cardinality of the auxiliary random variable can be obtained using arguments based on Caratheodory's theorem as described in \cite[Appendix~C]{Gam12}.

This concludes the proof of Theorem~\ref{thm:semidet_state_delay}.\hfill\IEEEQED

\section{Proof of Theorem~\ref{thm:mac}}\label{sec:mac}

The achievability scheme is more intricate than the previous sections due to the additional complications in the model, but builds on the same idea. Each encoder in the multiple access channel has an operation similar to the source encoder of the relay channels considered in the previous sections. The source encoder of the relay channels controlled the signal received at the relay by employing rate-splitting and superposition coding. This signal was used to choose a cooperation codeword for the next block. For the multiple access channel, each encoder controls the received signal at the other encoder in the same manner, so that at the end of a block, these two cribbed signals are known to both encoders, which are used to agree on a cooperation codeword for the next block. 

We point out the fact that it is crucial for both encoders to know both the cribbed signals at the end of a block, so that they can agree on a cooperation codeword for the next block. Encoder~1 knows $z_2^n$, because it receives this signal. Since the model assumes that the cribbing link is of the form $z_1(X_1,S_1)$, encoder~1 is able to control the $z_1^n$ signal received by encoder~2, and thus encoder~1 also knows $z_1^n$. Similarly, encoder~2 also knows $z_1^n$ and $z_2^n$. If $Z_1$ were assumed to be $z_1(X_1,X_2,S_1)$ or $z_1(X_1,X_2,S_1,S_2)$, then encoder~1 would not have knowledge of the received signal at encoder~2 due to the involvement of $X_2$ and $S_2$, and it would not be possible to employ the scheme. The reason we are able to assume that the received signal at the relay in the previous sections is $z(X,X_r,S)$ and not just $z(X,S)$ is that the relay has no message of its own, so $X_r$ in fact depends only on past signals transmitted by the source encoder, so the source encoder can still control the $z^n$ signal.

\subsection*{Proof:} 
Fix a pmf $p_U(u)p_{X_1|U,S_1}(x_1|u,s_1)p_{X_2|U,S_2}(x_2|u,s_2)$ and $\epsilon>0$. Split $R_1$ as $R'_1+R''_1$, with the message $M_1$ denoted accordingly as $(M'_1,M''_1)$, and similarly split $R_2$ as $R'_2+R''_2$, with the message $M_2$ denoted accordingly as $(M'_2,M''_2)$. Divide the total communication time into $B$ blocks, each of length $n$. 
In the achievability scheme proposed in \cite{Asn13} for the case of no state, $M'_1$ corresponds to the part of $M_1$ that is decoded by  encoder~2. As can be guessed based on the previous sections, this is not the case in the cooperative-bin-forward scheme presented below.


~

\subsubsection*{Codebook Generation}

~

For each block $b\in[1:B]$, a codebook is generated independently of the other blocks as follows:
\begin{itemize}
\item[-] \underline{\emph{Cooperation codewords}}\hfill\\
Generate $2^{n(\widetilde{R}_1+\widetilde{R}_2)}$ codewords $u_b^n(l_{1,b-1},l_{2,b-1})$, i.i.d. according to $p_U$, where $l_{1,b-1}\in[1:2^{n\widetilde{R}_1}]$ and $l_{2,b-1}\in[1:2^{n\widetilde{R}_2}]$. In the following, we will sometimes abbreviate $(l_{1,b-1},l_{2,b-1})$ by $l_{b-1}.$ 
\item[-] \underline{\emph{Cribbed codewords - I}}\hfill\\
For each $l_{b-1}$ and each $s_1\in\mc{S}_1$, generate a codebook of $2^{nR'_1}$ codewords. The $i$th symbol of such a codeword is chosen independently according to $p_{Z_1|U,S_1}(\cdot|u_{bi}(l_{b-1}),s_1)$. 
The result of this is that for each $l_{b-1}$, each $m'_{1,b}\in[1:2^{nR'_1}]$ and each $s_{1b}^n=(s_{1b1},s_{1b2},\dots ,s_{1bn})$, encoder~1 can form an effective codeword $z_{1b}^n(m'_{1,b} | l_{b-1},s_{1b}^n)$, whose $i$th symbol can be causally chosen as the $i$th symbol of the $m'_{1,b}$-th codeword from the codebook corresponding to $l_{b-1}$ and $s_{1bi}$. 
\item[-] \underline{\emph{Cribbed codewords - II}}\hfill\\
Similarly, for each $l_{b-1}$ and each $s_2\in\mc{S}_2$, generate a codebook of $2^{nR'_2}$ codewords. The $i$th symbol of such a codeword is chosen independently according to $p_{Z_2|U,S_2}(\cdot|u_{bi}(l_{b-1}),s_2)$. The result of this is that for each $l_{b-1}$, each $m'_{2,b}\in[1:2^{nR'_2}]$ and each $s_{2b}^n=(s_{2b1},s_{2b2},\dots ,s_{2bn})$, encoder~2 can form an effective codeword $z_{2b}^n(m'_{2,b} | l_{b-1},s_{2b}^n)$, whose $i$th symbol can be causally chosen as the $i$th symbol of the $m'_{2,b}$-th codeword from the codebook corresponding to $l_{b-1}$ and $s_{2bi}$. 
\item[-] \underline{\emph{Transmission codewords - I}}\hfill\\
For each $l_{b-1}$, each $m'_{1,b}\in[1:2^{nR'_1}]$ and each $s_1\in\mc{S}_1$, generate a codebook of $2^{nR''_1}$ codewords. The $i$th symbol of such a codeword is generated independently according to $p_{X_1|U,Z_1,S_1}(\cdot|u_{bi}(l_{b-1}),z_{1bi}(m'_{1,b}|l_{b-1},s_1),s_{1})$. The result of this construction is that for each $l_{b-1}$, each $m'_{1,b}\in[1:2^{nR'_1}]$, each $m''_{1,b}\in[1:2^{nR''_1}]$ and each $s_{1b}^n$, encoder~1 can form an effective codeword $x_{1b}^n(m''_{1,b}|l_{b-1},m'_{1,b},s_{1b}^n)$, whose $i$th symbol can be causally chosen as the $i$th symbol of the $m''_{1,b}$-th codeword from the codebook corresponding to $l_{b-1}$, $m'_{1,b}$ and $s_{1bi}$.
\item[-] \underline{\emph{Transmission codewords - II}}\hfill\\
Similarly, for each $l_{b-1}$, each $m'_{2,b}\in[1:2^{nR'_2}]$ and each $s_2\in\mc{S}_2$, generate a codebook of $2^{nR''_2}$ codewords. The $i$th symbol of such a codeword is generated independently according to $p_{X_2|U,Z_2,S_2}(\cdot|u_{bi}(l_{b-1}),z_{2bi}(m'_{2,b}|l_{b-1},s_2),s_{2})$. The result of this construction is that for each $l_{b-1}$, each $m'_{2,b}\in[1:2^{nR'_2}]$, each $m''_{2,b}\in[1:2^{nR''_2}]$ and each $s_{2b}^n$, encoder~2 can form an effective codeword $x_{2b}^n(m''_{2,b}|l_{b-1},m'_{2,b},s_{2b}^n)$, whose $i$th symbol can be causally chosen as the $i$th symbol of the $m''_{2,b}$-th codeword from the codebook corresponding to $l_{b-1}$, $m'_{2,b}$ and $s_{2bi}$.
\item[-] \underline{\emph{Binning}}\hfill\\
Finally, partition the set $\mathcal{Z}_{1}^n$ into $2^{n\widetilde{R}_1}$ bins, by choosing a bin for each $z_1^n$ independently and uniformly at random from $[1:2^{n\widetilde{R}_1}]$. Denote the chosen bin for $z_1^n$ by $\textsf{bin}_b(z_{1}^n)$. Similarly, partition the set $\mathcal{Z}_{2}^n$ into $2^{n\widetilde{R}_2}$ bins, by choosing a bin for each $z_2^n$ independently and uniformly at random from $[1:2^{n\widetilde{R}_2}]$. Denote the chosen bin for $z_2^n$ by $\textsf{bin}_b(z_{2}^n)$.
\end{itemize}

~

\subsubsection*{Encoding}

~

Henceforth, whenever convenient, we will abbreviate $(l_{1,b-1},l_{2,b-1})$ by $l_{b-1}$. Set $(l_{1,1}, l_{2,1}) = (1,1)$ and $$(m'_{1,B},m''_{1,B},m'_{2,B},m''_{2,B})=(1,1,1,1).$$ Since the message in the last block is fixed, the effective rate of communication will be $\left(\frac{B-1}{B}R_1, \frac{B-1}{B}R_2\right)$, which can be made as close as desired to $(R_1,R_2)$ by choosing a sufficiently large $B$. 
We now describe the encoding for block $b$. Assume both encoders have agreed upon some $l_{b-1} = (l_{1,b-1},l_{2,b-1})$ based on operations in previous blocks. Then, encoders encode messages $m_{1,b}$ and $m_{2,b}$ by $x^n_{1b}(m''_{1,b}|l_{b-1},m'_{1,b},s_{1b}^n)$ and $x^n_{2b}(m''_{2,b}|l_{b-1},m'_{2,b},s_{2b}^n)$ respectively. This operation is valid because it does not require noncausal knowledge of the state sequences, it can be done ``on the fly''. At the end of block $b$, both encoders have knowledge of the cribbed codewords $z_{1b}^n(m'_{1,b}|l_{b-1},s_{1b}^n)$ and $z_{2b}^n(m'_{2,b}|l_{b-1},s_{2b}^n)$. 
They set $$l_{1,b} = \textsf{bin}_b(z_{1b}^n), \quad\text{and}\quad l_{2,b}=\textsf{bin}_b(z_{2b}^n).$$

~

\subsubsection*{Decoding}

~

The decoder performs backward decoding. For each block $b\in\{B,B-1,B-2,\dots,2\}$, assuming that $l_b = (l_{1,b},l_{2,b})$ is known from previous operations:
\begin{itemize}
\item[(1)] The decoder first takes a pass through all $l_{b-1}=(l_{1,b-1},l_{2,b-1})$ and for each $l_{b-1}$, finds the unique $(m'_{1,b},m'_{2,b})$ such that 
$$\textsf{bin}_b(z_{1b}^n(m'_{1,b}|l_{b-1},s_{1b}^n)) = l_{1,b}\quad\text{ and }\quad\textsf{bin}_b(z_{2b}^n(m'_{2,b}|l_{b-1},s_{2b}^n)) = l_{2,b}.$$
Whenever a unique $(m'_{1,b},m'_{2,b})$ cannot be found for some $l_{b-1}$, the decoder chooses any $(m'_{1,b},m'_{2,b})$ arbitrarily. So after this operation, the decoder has chosen one $(m'_{1,b},m'_{2,b})$ for each $l_{b-1}$, given its knowledge of $(l_b,s_{1b}^n,s_{2b}^n)$. We will signify this explicitly by denoting the chosen messages as $\hat{m}'_{1,b}(l_{b-1},s_{1b}^n,l_{1,b})$ and $\hat{m}'_{2,b}(l_{b-1},s_{2b}^n,l_{2,b})$ respectively. 
\item[(2)] Now the decoder looks for the unique $(\hat{l}_{b-1},\hat{m}''_{1,b},\hat{m}''_{2,b})$ such that \eqref{eq:dec1} (appearing at the top of this page) is satisfied.
\end{itemize}

\begin{figure*}[!t]
\normalsize
\begin{IEEEeqnarray}{l}
\Big(u_b^n(\hat{l}_{b-1}),\,\,\,z_{1b}^n(\hat{m}'_{1,b}(\hat{l}_{b-1},s_{1b}^n,l_{1,b})\,|\,\hat{l}_{b-1},s_{1b}^n),\,\,\,z_{2b}^n(\hat{m}'_{2,b}(\hat{l}_{b-1},s_{2b}^n,l_{2,b}) \,| \,\hat{l}_{b-1},s_{2b}^n),\nonumber\\
\quad\quad\quad x_{1b}^n(\hat{m}''_{1,b}\,|\,\hat{l}_{b-1},\hat{m}'_{1,b}(\hat{l}_{b-1},s_{1b}^n,l_{1,b}),s_{1b}^n),\,\,\,x_{2b}^n(\hat{m}''_{2,b}\,|\,\hat{l}_{b-1},\hat{m}'_{2,b}(\hat{l}_{b-1},s_{2b}^n,l_{2,b}),s_{2b}^n),\,\,\,s_{1b}^n,s_{2b}^n,y_b^n\Big)\in\typ_eps \label{eq:dec1}
\end{IEEEeqnarray}
\hrulefill
\end{figure*}

~

\subsubsection*{Analysis of the Error Probability}\label{subsec:error_prob}

~

By symmetry, we can assume without loss of generality that the true messages and bin-indices corresponding to the current block are all 1, i.e. $${(L_{b-1},M'_{1,b},M'_{2,b},M''_{1,b},M''_{2,b})=(1,1,1,1,1)}.$$ We bound the probability of decoding error in block $b$ conditioned on successful decoding for blocks $\{B,B-1,\dots ,b+1\}$, averaged over the randomness in the messages and codebook generation. In particular, successful decoding in block $b+1$ means that $(L_{1,b},L_{2,b})$ has been decoded successfully, where we remind ourselves that
$$L_{1,b} = \textsf{Bin}_b(Z_{1b}^n(1|1,S_{1b}^n))\quad\text{and}\quad L_{2,b} = \textsf{Bin}_b(Z_{2b}^n(1|1,S_{2b}^n)).$$

An error occurs in block $b$ only if any of the following events occur:
\begin{itemize}
\item[$(a)$] $\hat{M}'_{1,b}(1,S_{1b}^n,L_{1,b})\neq 1$
\item[$(b)$] $\hat{M}'_{2,b}(1,S_{2b}^n,L_{2,b})\neq 1$
\item[$(c)$] $(\hat{L}_{b-1},\hat{M}''_{1,b},\hat{M}''_{2,b}) \neq (1,1,1)$ given $(\hat{M}'_{1,b}(1,S_{1b}^n,L_{1,b}),\hat{M}'_{2,b}(1,S_{2b}^n,L_{2,b}))= (1,1)$
\end{itemize}

We analyze each of the above three events in the following.

\subsubsection*{Event $(a)$: $\hat{M}'_{1,b}(1,S_{1b}^n,L_{1,b})\neq 1$} 
We have
\begin{IEEEeqnarray*}{L}
\text{Pr}\left(\hat{M}'_{1,b}(1,S_{1b}^n,L_{1,b})\neq 1\right)\\
\quad =\text{Pr}\left(\textsf{Bin}_b(Z_{1b}^n(m'_{1,b}|1,S_{1b}^n)) = L_{1,b} \text{ for some } m'_{1,b}>1\right)\\
\quad = \text{Pr}\left(\textsf{Bin}_b(Z_{1b}^n(m'_{1,b}|1,S_{1b}^n)) = \textsf{Bin}_b(Z_{1b}^n(1|1,S_{1b}^n)) \text{ for some } m'_{1,b}>1\right)\\
\quad \leq \sum_{m'_{1,b}>1} \text{Pr}\left(\textsf{Bin}_b(Z_{1b}^n(m'_{1,b}|1,S_{1b}^n)) = \textsf{Bin}_b(Z_{1b}^n(1|1,S_{1b}^n))\right)\\
\quad = \sum_{m'_{1,b}>1} \text{Pr}\left(\textsf{Bin}_b(Z_{1b}^n(m'_{1,b}|1,S_{1b}^n)) = \textsf{Bin}_b(Z_{1b}^n(1|1,S_{1b}^n)),\, Z_{1b}^n(m'_{1,b}|1,S_{1b}^n) =Z_{1b}^n(1|1,S_{1b}^n)\right)\\
\quad\quad +\> \sum_{m'_{1,b}>1} \text{Pr}\left(\textsf{Bin}_b(Z_{1b}^n(m'_{1,b}|1,S_{1b}^n)) = \textsf{Bin}_b(Z_{1b}^n(1|1,S_{1b}^n)), \, Z_{1b}^n(m'_{1,b}|1,S_{1b}^n) \neq Z_{1b}^n(1|1,S_{1b}^n)\right)\\
\quad \leq \sum_{m'_{1,b}>1} \text{Pr}\left( Z_{1b}^n(m'_{1,b}|1,S_{1b}^n) =Z_{1b}^n(1|1,S_{1b}^n)\right)\\
\quad\quad +\> \sum_{m'_{1,b}>1} \text{Pr}\left(\textsf{Bin}_b(Z_{1b}^n(m'_{1,b}|1,S_{1b}^n)) = \textsf{Bin}_b(Z_{1b}^n(1|1,S_{1b}^n)) \, | \, Z_{1b}^n(m'_{1,b}|1,S_{1b}^n) \neq Z_{1b}^n(1|1,S_{1b}^n)\right)\\
\quad \leq 2^{nR'_1}\cdot 2^{-n(H(Z_1|U,S_1)-\delta(\epsilon))} + 2^{nR'_1}\cdot 2^{-n\widetilde{R}_1},
\end{IEEEeqnarray*}
where we use $\delta(\epsilon)$ to denote any function of $\epsilon$ for which $\delta(\epsilon)\rightarrow 0$ as $\epsilon\rightarrow 0$. Hence, we get that
\begin{IEEEeqnarray*}{L}
\text{Pr}\left(\hat{M}'_{1,b}(1,S_{1b}^n,L_{1,b})\neq 1\right)\rightarrow 0 \text{ as } n\rightarrow\infty ,
\end{IEEEeqnarray*}
if the following two constraints are satisfied:
\begin{IEEEeqnarray*}{rCl}
R'_1 & < & \widetilde{R}_1,\\
R'_1 & < & H(Z_1|U,S_1) - \delta(\epsilon).
\end{IEEEeqnarray*}

\subsubsection*{Event $(b)$: $\hat{M}'_{2,b}(1,S_{2b}^n,L_{2,b})\neq 1$}
Similar to the previous subsection, we can conclude that 
\begin{IEEEeqnarray*}{L}
\text{Pr}\left(\hat{M}'_{2,b}(1,S_{2b}^n,L_{2,b})\neq 1\right) \rightarrow 0 \text{ as } n\rightarrow\infty ,
\end{IEEEeqnarray*}
 if the following two constraints are satisfied:
\begin{IEEEeqnarray*}{rCl}
R'_2 & < & \widetilde{R}_2,\\
R'_2 & < & H(Z_2|U,S_2) - \delta(\epsilon).
\end{IEEEeqnarray*}

\subsubsection*{Event $(c)$: $(\hat{L}_{b-1},\hat{M}''_{1,b},\hat{M}''_{2,b}) \neq (1,1,1)$ given $(\hat{M}'_{1,b}(1,S_{1b}^n,L_{1,b}),\hat{M}'_{2,b}(1,S_{2b}^n,L_{2,b}))= (1,1)$}

The probability of this event is upper bounded by 
\begin{IEEEeqnarray*}{CL}
&\text{Pr}\left(\text{Condition }\eqref{eq:dec1} \text{ is not satisfied by } (l_{b-1},m''_{1,b},m''_{2,b}) = (1,1,1) \,\, \big|\,\, (\hat{M}'_{1,b}(1,S_{1b}^n,L_{1,b}),\hat{M}'_{2,b}(1,S_{2b}^n,L_{2,b})) = (1,1)\right)\nonumber\\
+ & \text{Pr}\left(\text{Condition }\eqref{eq:dec1}\text{ is satisfied for some } (l_{b-1},m''_{1,b},m''_{2,b})\neq (1,1,1)\,\, \big|\,\, (\hat{M}'_{1,b}(1,S_{1b}^n,L_{1,b}),\hat{M}'_{2,b}(1,S_{2b}^n,L_{2,b})) = (1,1) \right).
\end{IEEEeqnarray*}

The first term 
goes to zero as $n\rightarrow\infty$ by the law of large numbers.

The second term can be handled by considering the following four different cases separately and applying the packing~lemma~\cite{Gam12} appropriately in each case.
\begin{itemize}
\item[-] $(\hat{L}_{b-1},\hat{M}''_{1,b},\hat{M}''_{2,b}) = (1,1,{>}1)$ given $(\hat{M}'_{1,b}(1,S_{1b}^n,L_{1,b}),\hat{M}'_{2,b}(1,S_{2b}^n,L_{2,b})) = (1,1)$
\item[-] $(\hat{L}_{b-1},\hat{M}''_{1,b},\hat{M}''_{2,b}) = (1,{>}1,1)$ given $(\hat{M}'_{1,b}(1,S_{1b}^n,L_{1,b}),\hat{M}'_{2,b}(1,S_{2b}^n,L_{2,b})) = (1,1)$
\item[-] $(\hat{L}_{b-1},\hat{M}''_{1,b},\hat{M}''_{2,b}) = (1,{>}1,{>}1)$ given $(\hat{M}'_{1,b}(1,S_{1b}^n,L_{1,b}),\hat{M}'_{2,b}(1,S_{2b}^n,L_{2,b})) = (1,1)$
\item[-] $(\hat{L}_{b-1},\hat{M}''_{1,b},\hat{M}''_{2,b}) = ({>}1,*,*)$
\end{itemize}

A standard application of the packing lemma gives us that the probability of each of the first three events goes to zero as $n\rightarrow\infty$ if the following constraints are respectively satisfied:
\begin{IEEEeqnarray*}{rCl}
R''_2 & < & I(X_2;Y|U,Z_2,X_1,S_1,S_2) - \delta(\epsilon),\\
R''_1 & < & I(X_1;Y|U,Z_1,X_2,S_1,S_2) - \delta(\epsilon),\\
R''_1 + R''_2 & < & I(X_1,X_2;Y|U,Z_1,Z_2,S_1,S_2) - \delta(\epsilon).
\end{IEEEeqnarray*}

Applying the packing lemma gives us the following condition for vanishing probability of the fourth event,
\begin{IEEEeqnarray*}{rCl}
\widetilde{R}_1 + \widetilde{R}_2 + R''_1 + R''_2 & < & I(U,Z_1,Z_2,X_1,X_2;Y|S_1,S_2) - \delta(\epsilon)\\
& = & I(X_1,X_2;Y|S_1,S_2) - \delta(\epsilon).
\end{IEEEeqnarray*}

Collecting all the constraints established so far, we have
\begin{IEEEeqnarray}{rCl}
R'_1 & < & \widetilde{R}_1,\label{eq:mac_first}\\
R'_1 & < & H(Z_1|U,S_1) - \delta(\epsilon),\\
R'_2 & < & \widetilde{R}_2,\\
R'_2 & < & H(Z_2|U,S_2) - \delta(\epsilon),\\
R''_2 & < & I(X_2;Y|U,Z_2,X_1,S_1,S_2) - \delta(\epsilon),\\
R''_1 & < & I(X_1;Y|U,Z_1,X_2,S_1,S_2) - \delta(\epsilon),\\
R''_1 + R''_2 & < & I(X_1,X_2;Y|U,Z_1,Z_2,S_1,S_2) - \delta(\epsilon),\\
\widetilde{R}_1 + \widetilde{R}_2 + R''_1 + R''_2 & < & I(X_1,X_2;Y|S_1,S_2) - \delta(\epsilon).\label{eq:mac_last}
\end{IEEEeqnarray}
Performing Fourier-Motzkin elimination of $\widetilde{R}_1$, $\widetilde{R}_2$, $R'_1$, $R'_2$, $R''_1$ and $R''_2$, and letting $n\rightarrow\infty$, $B\rightarrow\infty$ and $\epsilon\rightarrow 0$, we get that communication at arbitrarily small error probability is possible for the rates specified in Theorem~\ref{thm:mac}.

~

\subsubsection*{Converse}\label{subsec:conv}

~

The proof of the converse can be constructed by using similar arguments as in \cite{Asn13}. Note that we have by Fano's inequality the following condition for any reliable code:
$$H(M_1,M_2|Y^n,S_1^n,S_2^n) \leq n\epsilon_n,$$
where $\epsilon_n\rightarrow 0$ as $n\rightarrow\infty$. Define $U_i$ as $$U_i\triangleq (Z_1^{i-1},Z_2^{i-1},S_1^{i-1},S_2^{i-1}).$$An upper bound on $R_1$ is established by the following:
\begin{IEEEeqnarray*}{rCl}
nR_1 & = & H(M_1)\\
& \stackrel{(a)}{=} & H(M_1|M_2,S_1^n,S_2^n)\\
& \stackrel{(b)}{=} & H(M_1,Z_1^n|M_2,S_1^n,S_2^n)\\
& = & H(Z_1^n|M_2,S_1^n,S_2^n) + H(M_1|Z_1^n,M_2,S_1^n,S_2^n)\\
& \stackrel{(c)}{\leq} & H(Z_1^n|M_2,S_1^n,S_2^n) + I(M_1;Y^n|Z_1^n,M_2,S_1^n,S_2^n) + n\epsilon_n\\
& = & \sum_{i=1}^n H(Z_{1i}|Z_1^{i-1},M_2,S_1^n,S_2^n) + \sum_{i=1}^nI(M_1;Y_i|Y^{i-1},Z_1^n,M_2,S_1^n,S_2^n) + n\epsilon_n\\
& \stackrel{(d)}{=} & \sum_{i=1}^n H(Z_{1i}|Z_1^{i-1},M_2,S_1^n,S_2^n,Z_2^{i-1}) + \sum_{i=1}^nI(M_1;Y_i|Y^{i-1},Z_1^n,M_2,S_1^n,S_2^n,X_2^n) + n\epsilon_n\\
& \stackrel{(e)}{=} & \sum_{i=1}^n H(Z_{1i}|Z_1^{i-1},M_2,S_1^n,S_2^n,Z_2^{i-1}) + \sum_{i=1}^nI(M_1,X_{1i};Y_i|Y^{i-1},Z_1^n,M_2,S_1^n,S_2^n,X_2^n) + n\epsilon_n\\
& \stackrel{(f)}{\leq} & \sum_{i=1}^n H(Z_{1i}|Z_1^{i-1},S_{1}^i,S_2^{i-1},Z_2^{i-1}) + \sum_{i=1}^nI(X_{1i};Y_i|Z_1^{i},S_{1}^i,S_{2}^i,X_{2i},Z_{2}^{i-1}) + n\epsilon_n\\
& = & \sum_{i=1}^n H(Z_{1i}|U_i,S_{1i}) + \sum_{i=1}^nI(X_{1i};Y_i|U_i,Z_{1i},S_{1i},S_{2i},X_{2i}) + n\epsilon_n\\
& = & nH(Z_{1Q}|U_Q,S_{1Q},Q) + nI(X_{1Q};Y_Q|U_Q,Z_{1Q},S_{1Q},S_{2Q},X_{2Q},Q) + n\epsilon_n
\end{IEEEeqnarray*}
where
\begin{itemize}
\item[-] $Q$ is a random variable uniformly distributed on $[1:n]$, independent of other random variables,
\item[-] $(a)$ follows because $M_1$ is independent of $(M_2,S_1^n,S_2^n)$,
\item[-] $(b)$ follows since $Z_1^n$ is a function of $(M_1,M_2,S_1^n,S_2^n)$, 
\item[-] $(c)$ follows by Fano's inequality,
\item[-] $(d)$ follows since $(X_{2i},Z_{2i})$ is a function of $(M_2,S_{2}^n,Z_1^{i-1})$,
\item[-] $(e)$ follows since $X_{1i}$ is a function of $(M_1,S_1^n,X_{2}^{i-1})$,
\item[-] $(f)$ follows since (i) conditioning reduces entropy and (ii) conditioned on $(X_{1i},X_{2i},S_{1i},S_{2i})$, $Y_i$ is independent of $(M_1,M_2,X_1^n,X_2^n,S_1^n,S_2^n)$.
\end{itemize}

SImilarly, we get an upper bound on $R_2$:
\begin{IEEEeqnarray*}{rCl}
nR_2 & \leq & nH(Z_{2Q}|U_Q,S_{2Q},Q) + nI(X_{2Q};Y_Q|U_Q,Z_{2Q},S_{1Q},S_{2Q},X_{1Q},Q) + n\epsilon_n.
\end{IEEEeqnarray*}

Applying similar arguments to the sum rate, we get:
\begin{IEEEeqnarray*}{rCl}
n(R_1+R_2) & = & H(M_1,M_2)\\
& = & H(M_1,M_2|S_1^n,S_2^n)\\
& = & H(M_1,M_2,Z_1^n,Z_2^n|S_1^n,S_2^n)\\
& = & H(Z_1^n,Z_2^n|S_1^n,S_2^n) + H(M_1,M_2|Z_1^n,Z_2^n,S_1^n,S_2^n)\\
& \leq & \sum_{i=1}^n H(Z_{1i},Z_{2i}|U_i,S_{1i},S_{2i}) + H(M_1,M_2|Z_1^n,Z_2^n,S_1^n,S_2^n)\\
& \leq & \sum_{i=1}^n H(Z_{1i},Z_{2i}|U_i,S_{1i},S_{2i}) + I(M_1,M_2;Y^n|Z_1^n,Z_2^n,S_1^n,S_2^n) + n\epsilon_n\\
& \leq & \sum_{i=1}^n H(Z_{1i},Z_{2i}|U_i,S_{1i},S_{2i}) + \sum_{i=1}^nI(X_{1i},X_{2i};Y_i|U_i,Z_{1i},Z_{2i},S_{1i},S_{2i}) + n\epsilon_n\\
& = & nH(Z_{1Q},Z_{2Q}|U_Q,S_{1Q},S_{2Q},Q) + nI(X_{1Q},X_{2Q};Y_Q|U_Q,Z_{1Q},Z_{2Q},S_{1Q},S_{2Q},Q) + n\epsilon_n.
\end{IEEEeqnarray*}

Another upper bound on the sum rate can be easily established as follows:
\begin{IEEEeqnarray*}{rCl}
n(R_1+R_2) & = & H(M_1,M_2)\\
& = & H(M_1,M_2|S_1^n,S_2^n)\\
& \leq & I(M_1,M_2;Y^n|S_1^n,S_2^n) + n\epsilon_n\\
& \leq & \sum_{i=1}^n I(X_{1i},X_{2i};Y_i|S_{1i},S_{2i}) + n\epsilon_n\\
& = & nI(X_{1Q},X_{2Q};Y_Q|S_{1Q},S_{2Q},Q) + n\epsilon_n\\
& \leq & nI(X_{1Q},X_{2Q};Y_Q|S_{1Q},S_{2Q}) + n\epsilon_n.
\end{IEEEeqnarray*}

We note the following four conditions:
\begin{itemize}
\item[-] $(S_{1Q},S_{2Q})$ is independent of $(Q,U_Q)$, and has the pmf $p_{S_1,S_2}(s_1,s_2)$,
\item[-] $X_{1Q} - (Q,U_Q,S_{1Q}) - (Q,U_Q,S_{2Q}) - X_{2Q}$,
\item[-] $p_{Y_Q|U_Q,X_{1Q},X_{2Q},S_{1Q},S_{2Q}}(y|u,x_1,x_2,s_1,s_2)$ is equal to $p_{Y|X_{1},X_{2},S_{1},S_{2}}(y|x_1,x_2,s_1,s_2)$,
\item[-] $Z_{1Q}=z_1(X_{1Q},S_{1Q})$ and $Z_{2Q}=z_2(X_{2Q},S_{2Q})$.
\end{itemize}
The proof of the first condition follows because $(S_{1,i},S_{2,i})$ are generated independently of $U_i = (Z_1^{i-1},Z_2^{i-1},S_1^{i-1},S_2^{i-1})$ for all $1\leq i\leq n$, and due to the i.i.d. assumption on $(S_1,S_2)$. 
To prove that the second condition is satisfied, consider the following. For any $1\leq i\leq n$,\footnote{We drop subscripts denoting the random variables when analyzing the factorization to reduce the length of the expressions.}
\begin{IEEEeqnarray*}{lcl}
p(s_{1}^i,s_{2}^i,x_{1i},x_{2i},z_1^{i-1},z_2^{i-1}) \\
 = \sum_{m_1,m_2}p(m_1,m_2,s_1^i,s_2^i,x_{1i},x_{2i},z_1^{i-1},z_2^{i-1})\\ 
 = \sum_{m_1,m_2}\left[ p(m_1)p(m_2)p(s_1^{i},s_2^{i})p(x_{1i}|m_1,s_1^{i},z_2^{i-1})\left(\prod_{j=1}^{i-1}p(z_{1j}|m_1,s_1^j,z_2^{j-1})\right)\right.\\
 \quad\quad\quad\quad\quad\quad\quad\quad\quad\quad\quad\quad\times\,\left. p(x_{2i}|m_2,s_2^{i},z_1^{i-1})\left(\prod_{j=1}^{i-1}p(z_{2j}|m_2,s_2^j,z_1^{j-1})\right)\right]\\
 \footnote{The expressions from this line onwards, read in a literal manner, might seem strange for representing factorizations due to the presence of terms of the form $p(z_{1}^{i-1}|\cdot ,z_2^{i-1})p(z_2^{i-1}|\cdot ,z_1^{i-1})$, however these are valid factorizations, possibly containing redundant conditioning in some terms. The Markovity conclusions we draw from such a form are valid, because the factorization might at most contain redundant conditionings. An example might make the point clearer. Assume that a joint factorization is of the form $p(x,y)p(x',y')p(z|x,y')p(z'|x',y)$. If we marginalize by summing over $(x,x')$ first and then over $(z,z')$, we get an expression of the form $p(y|y')p(y'|y)$, while reversing the order of summation gives us $p(y)p(y')$. Thus, $Y$ and $Y'$ are independent, so the former expression is equal to the latter, though the former contains redundant conditioning.}= p(s_{1}^i,s_{2}^i)\sum_{m_1}p(m_1)p(x_{1i},z_1^{i-1}|m_1,s_1^{i},z_2^{i-1})\sum_{m_2} p(m_2)p(x_{2i},z_2^{i-1}|m_2,s_2^{i},z_1^{i-1})\\
=  p(s_{1}^i,s_{2}^i)p(x_{1i},z_1^{i-1}|s_{1}^i,z_2^{i-1})p(x_{2i},z_2^{i-1}|s_{2}^i,z_1^{i-1})\\
= p(s_{1}^i,s_{2}^i)p(x_{1i}|s_{1}^i,z_1^{i-1},z_2^{i-1})p(z_1^{i-1}|s_{1}^i,z_{2}^{i-1})p(x_{2i},z_2^{i-1}|s_{2}^i,z_1^{i-1}).
\end{IEEEeqnarray*}
The above factorization implies that 
\begin{IEEEeqnarray*}{RLL}
& X_{1i} - (S_{1}^i,Z_1^{i-1},Z_2^{i-1}) - (X_{2i},S_{2i}), & \text{ for all } 1\leq i\leq n\\
\Rightarrow \; & X_{1i} - (U_i,S_{1i}) - (X_{2i},S_{2i}), &\text{ for all } 1\leq i\leq n\\
\Rightarrow \; & X_{1i} - (U_i,S_{1i}) - (U_i,S_{2i},X_{2i}), &\text{ for all } 1\leq i\leq n\\
\Rightarrow \; & X_{1Q} - (U_Q,S_{1Q},Q) - (U_Q,S_{2Q},X_{2Q},Q).
\end{IEEEeqnarray*}
Similarly, we also have $$X_{2Q} - (U_Q,S_{2Q},Q) - (U_Q,S_{1Q},X_{1Q},Q).$$
These two Markov chains together imply the desired Markov chain $X_{1Q} - (U_Q,S_{1Q},Q) - (U_Q,S_{2Q},Q) - X_{2Q}$. The third condition follows by the definition of the auxiliary random variable $U_Q$ and because $Y_Q$ is the channel output when the inputs are $X_{1Q},X_{2Q},S_{1Q},S_{2Q}$. Similarly, the fourth condition is true because $Z_{1Q}$ and $Z_{2Q}$ are the cribbed signals due to $(X_{1Q},S_{1Q})$ and $(X_{2Q},S_{2Q})$ respectively.

So we can define random variables $U\triangleq (Q,U_Q)$, $X_1 \triangleq X_{1Q}$, $X_2 \triangleq X_{2Q}$, $S_1 \triangleq S_{1Q}$, $S_2 \triangleq S_{2Q}$, and $Y\triangleq Y_Q$, such that the joint pmf of these random variables has the factorization $$p_{S_1,S_2}(s_1,s_2)p_U(u)p_{X_1|U,S_1}(x_1|u,s_1)p_{X_2|U,S_2}(x_2|u,s_2)p_{Y|X_1,X_2,S_1,S_2}(y|x_1,x_2,s_1,s_2),$$ and $Z_1=z_1(X_1,S_1)$ and $Z_2=z_2(X_2,S_2)$. Noting that $\epsilon_n\rightarrow 0$ as $n\rightarrow\infty$, the four constraints on the rates that we have established become
\begin{equation*}
\begin{split}
R_1 & \leq I(X_1;Y|U,X_2,Z_1,S_1,S_2) + H(Z_1|U,S_1) ,\\
R_2 & \leq I(X_2;Y|U,X_1,Z_2,S_1,S_2) + H(Z_2|U,S_2),\\
R_1+R_2 & \leq I(X_1,X_2;Y|U,Z_1,Z_2,S_1,S_2)  + H(Z_1,Z_2|U,S_1,S_2),\\
R_1 + R_2 & \leq I(X_1,X_2;Y|S_1,S_2).
\end{split}
\end{equation*}
Thus, we get that the region stated in Theorem~\ref{thm:mac} is an outer bound to the achievable rate region. The bound on cardinality of the auxiliary random variable can be obtained using arguments based on Caratheodory's theorem as described in \cite[Appendix~C]{Gam12}. 

This concludes the proof of Theorem~\ref{thm:mac}.\hfill\IEEEQED

\section{Proof of Theorem~\ref{thm:mac_delay}}\label{sec:mac_delay}
The achievability part of this theorem builds on the cooperative-bin-forward scheme from the previous section by combining it with instantaneous relaying (a.k.a. codetrees or Shannon strategies). To avoid unnecessary repetition, we only provide the differences in the achievability part relative to that in the previous section. 

\subsection*{Proof:} 
Fix a pmf $p_U(u)p_{X_1|U,S_1}(x_1|u,s_1)p_{X_2|U,S_2,Z_1}(x_2|u,s_2,z_1)$ and $\epsilon>0$. Rate-splitting is performed as in the previous section. 

~

\subsubsection*{Codebook Generation}

~

The cooperation codewords and the codebooks used by Encoder 1 are generated in the same manner as the previous section. Encoder 2 generates codebooks by treating the causally observed $z_1$ symbol in the same manner as the causally observed $s_2$ symbol. More precisely, the codebooks constructed by Encoder 2 are described in the following two paragraphs.

For each $l_{b-1}$, each $s_2\in\mc{S}_2$ and each $z_1\in\mc{Z}_1$, generate a codebook of $2^{nR'_2}$ codewords. The $i$th symbol of such a codeword is chosen independently according to $p_{Z_2|U,S_2,Z_1}(\cdot|u_{bi}(l_{b-1}),s_2,z_1)$. The result of this is that for each $l_{b-1}$, each $m'_{2,b}\in[1:2^{nR'_2}]$, each $s_{2b}^n$ and each $z_{1b}^n$, encoder~2 can form an effective codeword $z_{2b}^n(m'_{2,b} | l_{b-1},s_{2b}^n,z_{1b}^n)$, whose $i$th symbol can be causally chosen as the $i$th symbol of the $m'_{2,b}$-th codeword from the codebook corresponding to $l_{b-1}$, $s_{2bi}$ and $z_{1,bi}$. 

For each $l_{b-1}$, each $m'_{2,b}\in[1:2^{nR'_2}]$, each $s_2\in\mc{S}_2$ and each $z_{1}\in\mc{Z}_1$, generate a codebook of $2^{nR''_2}$ codewords. The $i$th symbol of such a codeword is generated independently according to $p_{X_2|U,Z_2,S_2,Z_1}(\cdot|u_{bi}(l_{b-1}),z_{2bi}(m'_{2,b}|l_{b-1},s_2,z_1),s_{2},z_1)$. The result of this construction is that for each $l_{b-1}$, each $m'_{2,b}\in[1:2^{nR'_2}]$, each $m''_{2,b}\in[1:2^{nR''_2}]$, each $s_{2b}^n$ and each $z_{1b}^n$, encoder~2 can form an effective codeword $x_{2b}^n(m''_{2,b}|l_{b-1},m'_{2,b},s_{2b}^n,z_{1b}^n)$, whose $i$th symbol can be causally chosen as the $i$th symbol of the $m''_{2,b}$-th codeword from the codebook corresponding to $l_{b-1}$, $m'_{2,b}$, $s_{2bi}$ and $z_{1,bi}$.

The binning is performed as in the previous section.

~

\subsubsection*{Encoding}

~

The encoding at Encoder~1 is identical to that in the previous section. Encoder~2  transmits $x_{2b}^n(m''_{2,b}|l_{b-1},m'_{2,b},s_{2b}^n,z_{1b}^n)$ which can be chosen as described above.

~

\subsubsection*{Decoding}

~

The decoder performs backward decoding over the blocks, where it performs two steps as in the previous section, with the first step changed to the following.

The decoder first takes a pass through all $l_{b-1}=(l_{1,b-1},l_{2,b-1})$ and for each $l_{b-1}$, finds the unique $(m'_{1,b},m'_{2,b})$ such that 
$$\textsf{bin}_b(z_{1b}^n(m'_{1,b}|l_{b-1},s_{1b}^n)) = l_{1,b}\quad\text{ and }\quad\textsf{bin}_b(z_{2b}^n(m'_{2,b}|l_{b-1},s_{2b}^n,z_{1b}^n(m'_{1,b}|l_{b-1},s_{1b}^n))) = l_{2,b}.$$

~

\subsubsection*{Probability of Error}

~

In the previous section, we obtained the conditions \eqref{eq:mac_first}-\eqref{eq:mac_last} for vanishing probability of error. The only difference now is that the fourth condition is replaced by 
$$R'_2 < H(Z_2|U,S_2,Z_1) - \delta(\epsilon).$$
This is obtained by analyzing the probability of event $(b)$ conditioned on the complement of event $(a)$. The other conditions remain the same. Performing Fourier-Motzkin elimination of $\widetilde{R}_1$, $\widetilde{R}_2$, $R'_1$, $R'_2$, $R''_1$ and $R''_2$, and letting $n\rightarrow\infty$, $B\rightarrow\infty$ and $\epsilon\rightarrow 0$, we get that communication at arbitrarily small error probability is possible for the rates specified in Theorem~\ref{thm:mac_delay}.

~

\subsubsection*{Converse}

~

The only difference in the converse compares to that of the previous section is that we need to show a different bound on $R_2$ and we need to prove the factorization of the pmf. The bound on $R_1$ and the two bounds on the sum rate $R_1+R_2$ are the same and require no changes in the arguments. 

The new bound on $R_2$ can be shown by following the same line of arguments with minor changes. We provide the chain of inequalities below for completeness. The auxiliary random variable $U_i$ appearing below is defined to be $(Z_1^{i-1},Z_2^{i-1},S_1^{i-1},S_2^{i-1})$.

\begin{IEEEeqnarray*}{rCl}
nR_2 & = & H(M_2)\\
& \stackrel{(a)}{=} & H(M_2|M_1,S_1^n,S_2^n)\\
& \stackrel{(b)}{=} & H(M_2,Z_2^n|M_1,S_1^n,S_2^n)\\
& = & H(Z_2^n|M_1,S_1^n,S_2^n) + H(M_2|Z_2^n,M_1,S_1^n,S_2^n)\\
& \stackrel{(c)}{\leq} & H(Z_2^n|M_1,S_1^n,S_2^n) + I(M_2;Y^n|Z_2^n,M_1,S_1^n,S_2^n) + n\epsilon_n\\
& = & \sum_{i=1}^n H(Z_{2i}|Z_2^{i-1},M_1,S_1^n,S_2^n) + \sum_{i=1}^nI(M_2;Y_i|Y^{i-1},Z_2^n,M_1,S_1^n,S_2^n) + n\epsilon_n\\
& \stackrel{(d)}{=} & \sum_{i=1}^n H(Z_{2i}|Z_2^{i-1},M_1,S_1^n,S_2^n,Z_1^{i}) + \sum_{i=1}^nI(M_2;Y_i|Y^{i-1},Z_2^n,M_1,S_1^n,S_2^n,X_1^n) + n\epsilon_n\\
& \stackrel{(e)}{=} & \sum_{i=1}^n H(Z_{2i}|Z_2^{i-1},M_1,S_1^n,S_2^n,Z_1^{i}) + \sum_{i=1}^nI(M_2,X_{2i};Y_i|Y^{i-1},Z_2^n,M_1,S_1^n,S_2^n,X_1^n) + n\epsilon_n\\
& \stackrel{(f)}{\leq} & \sum_{i=1}^n H(Z_{2i}|Z_2^{i-1},S_{2}^i,S_1^{i-1},Z_1^{i}) + \sum_{i=1}^nI(X_{2i};Y_i|Z_2^{i},S_{1}^i,S_{2}^i,X_{1i},Z_{1}^{i-1}) + n\epsilon_n\\
& = & \sum_{i=1}^n H(Z_{2i}|U_i,S_{2i},Z_{1i}) + \sum_{i=1}^nI(X_{2i};Y_i|U_i,Z_{2i},S_{1i},S_{2i},X_{1i}) + n\epsilon_n\\
& = & nH(Z_{2Q}|U_Q,S_{2Q},Z_{1Q},Q) + nI(X_{2Q};Y_Q|U_Q,Z_{2Q},S_{1Q},S_{2Q},X_{1Q},Q) + n\epsilon_n
\end{IEEEeqnarray*}
where
\begin{itemize}
\item[-] $Q$ is a random variable uniformly distributed on $[1:n]$, independent of other random variables,
\item[-] $(a)$ follows because $M_2$ is independent of $(M_1,S_1^n,S_2^n)$,
\item[-] $(b)$ follows since $Z_2^n$ is a function of $(M_1,M_2,S_1^n,S_2^n)$, 
\item[-] $(c)$ follows by Fano's inequality,
\item[-] $(d)$ follows since $(X_{1i},Z_{1i})$ is a function of $(M_1,S_{1}^n,Z_2^{i-1})$,
\item[-] $(e)$ follows since $X_{2i}$ is a function of $(M_2,S_2^n,X_{1}^{i})$,
\item[-] $(f)$ follows since (i) conditioning reduces entropy and (ii) conditioned on $(X_{1i},X_{2i},S_{1i},S_{2i})$, $Y_i$ is independent of $(M_1,M_2,X_1^n,X_2^n,S_1^n,S_2^n)$.
\end{itemize}

Regarding the joint pmf, we note the following conditions
\begin{itemize}
\item[-] $(S_{1Q},S_{2Q})$ is independent of $(Q,U_Q)$, and has the pmf $p_{S_1,S_2}(s_1,s_1)$,
\item[-] $X_{1Q} - (Q,U_Q,S_{1Q}) - S_{2Q}$,
\item[-] $X_{2Q} - (Q,U_Q,S_{2Q},Z_{1Q}) - (X_{1Q},S_{1Q})$,
\item[-] $p_{Y_Q|U_Q,X_{1Q},X_{2Q},S_{1Q},S_{2Q}}(y|u,x_1,x_2,s_1,s_2)$ is equal to $p_{Y|X_{1},X_{2},S_{1},S_{2}}(y|x_1,x_2,s_1,s_2)$,
\item[-] $Z_{1Q}=z_1(X_{1Q},S_{1Q})$ and $Z_{2Q}=z_2(X_{2Q},S_{2Q})$.
\end{itemize}
The first, fourth and fifth conditions do not need new arguments. The second and third condition can be proved as follows. For any $1\leq i\leq n$,
\begin{IEEEeqnarray*}{lcl}
p(s_{1}^i,s_{2}^i,x_{1i},z_1^{i-1},z_2^{i-1}) \\
= \sum_{m_1,m_2} p(m_1,m_2,s_{1}^i,s_{2}^i,x_{1i},z_1^{i-1},z_2^{i-1}) \\
= \sum_{m_1,m_2} p(m_1)p(m_2)p(s_1^i,s_2^i)p(x_{1i}|m_1,s_1^i,z_2^{i-1})\prod_{j=1}^{i-1}p(z_{1j}|m_1,s_1^j,z_{2}^{j-1})\prod_{j=1}^{i-1}p(z_{2j}|m_2,s_2^{j},z_1^{j})\\
= p(s_1^i,s_2^i)p(x_{1i},z_1^{i-1}|s_1^i,z_2^{i-1})p(z_{2}^{i-1}|s_2^{i},z_1^{i}).
\end{IEEEeqnarray*}
This implies $X_{1Q} - (U_Q,S_{1Q},Q) - S_{2Q}.$ For the third condition, we have for any $1\leq i\leq n$,
\begin{IEEEeqnarray*}{lcl}
p(s_{1}^i,s_{2}^i,x_{1i},x_{2i},z_1^{i-1},z_{1i},z_2^{i-1}) \\
 = \sum_{m_1,m_2}p(m_1,m_2,s_1^i,s_2^i,x_{1i},x_{2i},z_1^{i-1},z_{1i},z_2^{i-1})\\ 
 = \sum_{m_1,m_2}\left[ p(m_1)p(m_2)p(s_1^{i},s_2^{i})p(x_{1i}|m_1,s_1^{i},z_2^{i-1})\left(\prod_{j=1}^{i}p(z_{1j}|m_1,s_1^j,z_2^{j-1})\right)\right.\\
 \quad\quad\quad\quad\quad\quad\quad\quad\quad\quad\quad\quad\times\,\left. p(x_{2i}|m_2,s_2^{i},z_1^{i})\left(\prod_{j=1}^{i-1}p(z_{2j}|m_2,s_2^j,z_1^{j})\right)\right]\\
 = p(s_{1}^i,s_{2}^i)\sum_{m_1}p(m_1)p(x_{1i},z_1^{i}|m_1,s_1^{i},z_2^{i-1})\sum_{m_2} p(m_2)p(x_{2i},z_2^{i-1}|m_2,s_2^{i},z_1^{i})\\
=  p(s_{1}^i,s_{2}^i)p(x_{1i},z_1^{i}|s_{1}^i,z_2^{i-1})p(x_{2i},z_2^{i-1}|s_{2}^i,z_1^{i})
\end{IEEEeqnarray*}
The above factorization implies that 
\begin{IEEEeqnarray*}{RLL}
& X_{2i} - (S_{2}^i,Z_1^{i},Z_2^{i-1}) - (X_{1i},S_{1i}), & \text{ for all } 1\leq i\leq n\\
\Rightarrow \; & X_{2i} - (U_i,S_{2i},Z_{1i}) - (X_{1i},S_{1i}), &\text{ for all } 1\leq i\leq n\\
\Rightarrow \; & X_{2Q} - (U_Q,S_{2Q},Z_{1Q},Q) - (X_{1Q},S_{1Q}).
\end{IEEEeqnarray*}

So we can define random variables $U\triangleq (Q,U_Q)$, $X_1 \triangleq X_{1Q}$, $X_2 \triangleq X_{2Q}$, $S_1 \triangleq S_{1Q}$, $S_2 \triangleq S_{2Q}$, and $Y\triangleq Y_Q$, such that the joint pmf of these random variables has the factorization $$p_{S_1,S_2}(s_1,s_2)p_U(u)p_{X_1|U,S_1}(x_1|u,s_1)p_{X_2|U,S_2,Z_1}(x_2|u,s_2,z_1)p_{Y|X_1,X_2,S_1,S_2}(y|x_1,x_2,s_1,s_2),$$ and $Z_1=z_1(X_1,S_1)$ and $Z_2=z_2(X_2,S_2)$. Noting that $\epsilon_n\rightarrow 0$ as $n\rightarrow\infty$, the four constraints on the rates that we have established become
\begin{equation*}
\begin{split}
R_1 & \leq I(X_1;Y|U,X_2,Z_1,S_1,S_2) + H(Z_1|U,S_1) ,\\
R_2 & \leq I(X_2;Y|U,X_1,Z_2,S_1,S_2) + H(Z_2|U,S_2,Z_1),\\
R_1+R_2 & \leq I(X_1,X_2;Y|U,Z_1,Z_2,S_1,S_2)  + H(Z_1,Z_2|U,S_1,S_2),\\
R_1 + R_2 & \leq I(X_1,X_2;Y|S_1,S_2).
\end{split}
\end{equation*}
Thus, we get that the region stated in Theorem~\ref{thm:mac_delay} is an outer bound to the achievable rate region. The bound on cardinality of the auxiliary random variable can be obtained using arguments based on Caratheodory's theorem as described in \cite[Appendix~C]{Gam12}. 

This concludes the proof of Theorem~\ref{thm:mac_delay}.\hfill\IEEEQED

\section{Concluding Remarks and Some Open Problems}\label{sec:conc}
We presented the cooperative-bin-forward scheme and showed that it achieves the capacity region in a variety of semideterministic setups. While partial-decode-forward has been the scheme of interest in semideterministic setups, we demonstrated the strict advantages of cooperative-bin-forward by considering state-dependent setups, where partial-decode-forward cannot be applied, but cooperative-bin-forward is optimal.

A number of interesting questions remain. Most importantly, how can the cooperative-bin-forward scheme be extended to, e.g. the model in Figure~\ref{fig:state_semidet_relay}, when the source-relay link is not deterministic, but a general noisy link? Cooperative-bin-forward was developed in this paper as an alternative to partial-decode-forward. However, the latter has an advantage of extending naturally to the noisy case due to the decoding operation at the relay. 
The crucial high-level ingredient for establishing cooperation that is used by both schemes is that different nodes agree on some information. In partial-decode-forward, the agreement is established via a decoding operation at the relay. In cooperative-bin-forward, the decoding operation was removed and the agreement was established by exploiting the deterministic components in the models. To ensure some kind of agreement between nodes in the general noisy case without using a decoding operation, the similarity between the operations of binning and compression suggest an approach. Note that binning is a form of compression, so a natural extension of cooperative-bin-forward might involve a compression operation at the relay, where part of the compression can be reconstructed at the source, thus enabling some cooperation between the source and the relay.

Another interesting question is that of designing optimal achievability schemes for all the state-dependent setups considered in this paper when the state is known only to the source encoders, causally or strictly causally. Finally, the semideterministic relay channel with two state components, one known to the source and the other to the relay, with an uninformed destination, is also an interesting open question. 


\section*{Acknowledgments}
The authors gratefully acknowledge helpful discussions with Xiugang Wu.

\bibliographystyle{IEEEtran}
\bibliography{IEEEfull,mac}

\end{document}